\documentclass[12pt]{article}
\usepackage{paper2e}
\usepackage{epsfig}
\usepackage{amssymb}

\begin{document}
\begin{titlepage}

\preprint{UTAP-547, RESCEU-40/05}

\title{Dynamical Stability of Six-dimensional Warped Flux Compactification}

\author{Hiroyuki Yoshiguchi$^{\rm a}$, 
Shinji Mukohyama$^{\rm a,b}$,\\
Yuuiti Sendouda$^{\rm a}$,
Shunichiro Kinoshita$^{\rm a}$
}

\address{$^{\rm a}$Department of Physics\\
The University of Tokyo, Tokyo 113-0033, Japan}

\address{$^{\rm b}$Research Center for the Early Universe\\
The University of Tokyo, Tokyo 113-0033, Japan}

\begin{abstract}
We show the dynamical stability of a six-dimensional braneworld solution
with warped flux compactification recently found by the authors. 
We consider linear perturbations around this background spacetime, 
assuming the axisymmetry in the extra dimensions. The perturbations
are expanded by scalar-, vector- and tensor-type harmonics of the
four-dimensional Minkoswki spacetime and we analyze each type
separately. 
It is found that there is no unstable mode in each sector and that  
there are zero modes only in the tensor sector, corresponding to the
four-dimensional gravitons.
We also obtain the first few Kaluza-Klein modes in each sector.
\end{abstract}

\end{titlepage}

\section{Introduction}

It is being established that there are two major accelerating, or quasi 
de Sitter, phases in the history of our universe. One is inflation in
the early universe and the other is the current phase of accelerating
expansion. This picture is supported by observations such as cosmic
microwave background and supernovae~\cite{Perlmutter,Schmidt,Riess}, and
thus we are almost sure that there are and were such accelerating
phases. However, we essentially do not know what causes those phases,
i.e. we do not yet know what inflaton and dark energy are from the
viewpoint of fundamental physics.

In string theory it had been thought rather difficult to construct a
four-dimensional de Sitter universe with stabilized moduli until the
construction by Kachru, Kallosh, Linde and Trivedi
(KKLT)~\cite{KKLT}. (See \cite{EGQ,BKQ} for followup proposals.) They 
evaded previously known no-go theorems by putting anti-D-branes at the
bottom of a warped throat after stabilizing all moduli. Since the shape
of the warped, compact extra dimensions are stabilized by fluxes, this
set-up is often called warped flux compactification. This set-up
provides a number of possible applications to
cosmology~\cite{KKLMMT,BBCEGKLQ,InflationKKLTsetup,Polchinski,Mukohyama2005}.

In our previous paper~\cite{paper1} we pointed out that brane gravity in
the warped flux compactification is somehow similar to that in the first
Randall-Sundrum (RS1) scenario~\cite{RS1} with radion
stabilization~\cite{GW}. (See
refs.~\cite{Tanaka-Montes,CGK,Mukohyama-Kofman,Mukohyama-HDterm} for
brane gravity in the RS1 scenario with radion stabilization.) The
evolution of matter on the brane changes the bulk geometry not only near
the brane but possibly everywhere in the whole extra
dimensions. Provided that all moduli are stabilized, the bulk geometry
should quickly settle to a configuration which is determined by the 
boundary condition, i.e. the brane source(s), values of conserved
quantities and the regularity of the other region of the extra
dimensions. As a consequence of the change of the bulk geometry, the
induced geometry on the brane responds to the evolution of the matter
source on the brane. The four-dimensional Einstein theory is recovered as
a rather indirect and subtle relation between the matter source on the
brane and the response of the induced geometry. To support this picture,
we considered a simplified situation in which we can see the recovery of
the four-dimensional Einstein theory in the warped flux
compactification. In particular, we found an exact solution representing
a six-dimensional brane world with warped flux compactification, 
including a warped geometry, compactification, a magnetic flux, and one
or two $3$-brane(s). (See
refs.~\cite{LW,Carroll-Guica,Garriga-Porrati,Navarro,Vinet-Cline,CDGV}
for other models of six-dimensional brane world.)

The purpose of this paper is to show the stability of the
six-dimensional exact solution found in the previous
paper~\cite{paper1}. For simplicity we set the four-dimensional
cosmological constant to zero and assume axisymmetry in the extra
dimensions. We expand linear perturbations by scalar-, vector- and
tensor-type harmonics of the four-dimensional Minkowski spacetime and
analyze each type separately. 
Linear perturbations in six-dimensional models are also considered in
\cite{RT,LP}.
We shall show that there is no unstable
mode in each sector and that there are zero modes only in the tensor
sector, corresponding to the four-dimensional gravitons. We also obtain
the first few Kaluza-Klein modes in each sector.

The rest of this paper is organized as follows.
In section~\ref{background}, we briefly review our six-dimensional
brane world model. We then consider linear perturbations with
axisymmetry to show the stability of our exact solution in
section~\ref{stability}. We numerically solve the perturbed Einstein
equations and Maxwell equations for each type of perturbations, and then
show that there is no unstable mode. In section~\ref{summary}, we
summarize the main results and discuss them.

\section{6D Warped Flux Compactification}
\label{background}

In this section, we briefly review our six-dimensional braneworld model,
which captures essential features of the warped flux compactification,
that is, warped geometry, magnetic flux of an antisymmetric field along
the extra dimensions, and branes.
The model is simple enough to make it possible for us to analyze gravity
on the brane from higher dimensional point of view.
Our start point is the six-dimensional Einstein-Maxwell action,
%
%
\begin{equation}
I_6 = \frac{M_6^4}{2} \int d^6 x \sqrt{-g} \biggl( R -2\Lambda_6
 -\frac{1}{2} F^{MN}F_{MN} \biggr),
\end{equation}
where $M_6$ is the six-dimensional reduced Planck mass, $\Lambda_6$ is the
bulk cosmological constant, and $F_{MN}=\partial_M A_N - \partial_N A_M$
is the field strength associated with the $U(1)$ gauge field $A_M$.

The bulk solution considered as the background in this study is 
%
%
\begin{eqnarray}
&&d s_6^2 = r^2 \eta_{\mu\nu} dx^\mu dx^\nu + \frac{dr^2}{f(r)} + f(r)
 d\phi^2,\\
&&A_M dx^M = A(r) d\phi,
\end{eqnarray}
where
%
%
\begin{eqnarray}
&&f(r) = -\frac{\Lambda_6}{10} r^2 -\frac{\mu_b}{r^3} -\frac{b^2}{12r^6},
\label{eq_f}\\
&&A(r) = \frac{b}{3r^3}.
\end{eqnarray}
This solution corresponds to the $\Lambda_{4\pm}\to 0$ limit of the more
general solution found in \cite{paper1}, where $\Lambda_{4\pm}$ is the 
four-dimensional cosmological constant on the branes. 
(See Appendix A.2. of \cite{paper1} for this limit.)
As explained in \cite{paper1}, this solution is related by a double Wick
rotation to a topological black hole with the horizon topology ${\bf R}^4$.

Next, we consider embedding of one or two 3-brane sources.
As is well known, an object with codimension 2 induces a deficit angle 
around it as
%
%
\begin{equation}
\delta_\pm = \frac{\sigma_\pm}{M_6^4},
 \label{eqn:deficit-angle-formula}
\end{equation}
where $\delta_\pm$ is the deficit angle due to the tension $\sigma_\pm$
of the brane.
We should note that this formula is valid under the axisymmetry if
radial stress is much smaller than energy density~\cite{FIU}.

Hereafter, we assume that the function $f(r)$ given by Eq.(\ref{eq_f})
have two positive roots $r=r_\pm$ ($0 < r_- <r_+$) for $ f(r)=0 $ and is positive
between them ($r_- < r <r_+$).
This requires that the six-dimensional cosmological constant $\Lambda_6$
be positive, and thus we assume $\Lambda_6>0$ throughout the paper. 
Since $f$ vanishes at $r_\pm$, $r = r_\pm$ defines surfaces of
codimension 2.
Thus we can put the 3-branes at $r_\pm$.
The periods of the angular coordinate $\Delta \phi$ calculated at
$r_\pm$ must coincide, thus we get
%
%
\begin{equation}
\Delta \phi = \frac{2\pi - \delta_+}{\kappa_+} = \frac{2\pi -
 \delta_-}{\kappa_-},
\label{period}
\end{equation}
where $ \delta_\pm $ are the deficit angles,
%
%
\begin{equation}
\kappa_\pm \equiv \mp \frac{1}{2} f'(r_\pm),
\label{kappa_def}
\end{equation}
and a prime denotes derivative with respect to $r$.
This is rewritten as
%
%
\begin{equation}
\frac{2\pi - \delta_+}{2\pi - \delta_-} = \frac{\kappa_+}{\kappa_-},
\label{bc_phi}
\end{equation}
which can be regarded as a boundary condition since the l.h.s is
specified by the brane sources and the r.h.s. can be written in terms of
the bulk parameters $\mu_b$ and $b$.

The background geometry can be specified by
the three parameters, the tensions $\sigma_\pm$ of the brane at $r_\pm$
and the magnetic flux $\Phi$ in the bulk.
The magnetic flux $\Phi$ is given by the equation
%
%
\begin{equation}
\frac{\Phi}{\sqrt{(2\pi - \delta_+)(2\pi - \delta_-)}} =
 -\frac{b}{3\sqrt{\kappa_+ \kappa_-} L}
\biggl( \frac{1}{r_-^3} -  \frac{1}{r_+^3}\biggr),
\label{bc_flux}
\end{equation}
where $L \equiv \sqrt{10/\Lambda_6}$ and the r.h.s is written only by
$\mu_b$ and $b$.
If we specify $\sigma_\pm$ and $\Phi$, the left hand sides of
Eq.(\ref{bc_phi}) and Eq.(\ref{bc_flux}) are determined.
On the other hand, the right hand sides of these equations are written
by $\mu_b$ and $b$.
Thus we can solve (\ref{bc_phi}) and (\ref{bc_flux}) w.r.t. $\mu_b$ and
$b$. We can also determine $\Delta \phi$ by Eq.(\ref{period}).

Before going into details, we make some general remarks on the
background solution. Our exact braneworld solution captures some
essential features of the warped flux compactification, including warped
geometry, compactification, moduli stabilization, flux and
branes. However, since the higher-dimensional cosmological constant
$\Lambda_6$ is assumed to be positive, we have to confess that this
model of warped flux compactification does not completely mimic the
KKLT construction, where the $10$-dimensional cosmological constant is
not positive but zero. Nonetheless, our model is useful in the sense
that it provides a testing ground on which brane gravity with warped
flux compactification can be analyzed from higher dimensional
viewpoints. Note that many cosmological considerations in the KKLT setup
are based on the implicit assumption that $4$-dimensional Einstein
gravity should be recovered at low energy and that it is very important
to see whether this assumption is viable or not from higher dimensional 
viewpoint. As discussed in \cite{paper1}, the recovery of the
$4$-dimensional Einstein gravity in warped flux compactification is not
as simple as would be expected from the $4$-dimensional effective
theory. The $4$-dimensional gravity is 
recovered in a rather subtle way as a consequence of bulk dynamics, and
the exact solutions in our model were useful to see how this picture
works explicitly.

Finally, we explain the limit $\alpha\equiv r_-/r_+ \to 1$ of the bulk
geometry. In this limit, the coordinate distance $r_+-r_-$ between the
two branes vanishes, and thus the bulk geometry appears to
collapse. However, the proper distance between $r=r_-$ and $r_+$ does
not vanish and the geometry of extra dimensions remains regular. 
This can be explicitly shown by a coordinate transformation:
%
\begin{eqnarray}
\bar r &=& \frac{2r - \left(r_+ + r_-\right)}{r_+-r_-}
~~~\left( -1\leq \bar r\leq 1 \right),
\label{rbar}
\\
\varphi &=& \Lambda_6 \left(r_+ - r_- \right) \phi
\label{varphi_def}
\end{eqnarray}
Even if we take $\alpha \to 1$, the domain of $\bar r$ remains finite
while that of $r$ vanishes.
With this new coordinate system, the metric of the extra dimensions
becomes 
%
%
\begin{equation}
\frac{dr^2}{f(r)} + f(r) d\phi^2 = \frac{d\bar r^2}{4\bar f}
 + \frac{\bar f}{\Lambda_6^2} d\varphi^2,
\label{metric_rbar}
\end{equation}
where
%
\begin{eqnarray}
\bar f \left(\bar r\right) &=& 
\frac{f\left(r\right)}{r_+ r_-
\left(\alpha^{-1/2}-\alpha^{1/2}\right)^2}
\nonumber\\
&=&\frac{1}{\beta_-^2}\biggl[ -\frac{\gamma_2}{\gamma_1}
\left(\frac{2}{\beta_- \bar r + \beta_+}\right)^6
+\frac{\beta_+ (\alpha^{-1}+\alpha)(\alpha^{-2}+\alpha^2)}{\gamma_1}
\left(\frac{2}{\beta_- \bar r + \beta_+}\right)^3
\nonumber\\
&&~~~~~~~~~~
-\left(\frac{\beta_- \bar r + \beta_+}{2}\right)^2 \biggr],
\end{eqnarray}
and we defined
%
\begin{eqnarray}
\beta_\pm &\equiv& \alpha^{-1/2} \pm \alpha^{1/2},
\\
\gamma_n &\equiv& \sum_{i=0}^{2n}\alpha^{i-n}.
\end{eqnarray}
The function $\bar f$ includes only the background parameter $\alpha$,
and is defined so as to remain finite in the limit of $\alpha\to 1$. 
Indeed, $\bar{f}$ in this limit is as simple as 
%
\begin{equation}
\bar f = \frac{\Lambda_6}{2} \left(1-\bar r^2\right).
\label{fbar_limit}
\end{equation}
Thus, the metric (\ref{metric_rbar}) becomes that of a round sphere of
radius $1/\sqrt{2\Lambda_6}$ in this limit.

As we mentioned, the coordinate $\bar r$ runs over the finite interval
$\left[-1,1\right]$.
On the other hand, the period of the coordinate $\varphi$ appears to
collapse since the coefficient $\left(r_+ - r_- \right)$ in the
definition (\ref{varphi_def}) vanishes in the $\alpha\to 1$ limit.
Actually, this is not the case.
The ``surface gravity'' $\kappa_\pm$ defined in Eq.(\ref{kappa_def}) is
written in terms of $\bar f$ as
%
\begin{equation}
\kappa_\pm = \mp \left(r_+ - r_-\right) \partial_{\bar r}\bar f
\left(\pm 1\right).
\end{equation}
Therefore, the new coordinate $\varphi$ has the period
%
\begin{equation}
\Delta \varphi = \Lambda_6 \left(r_+ - r_-\right) \Delta\phi
=\mp \frac{\Lambda_6}{\partial_{\bar r}\bar f
\left(\pm 1\right)}\left(2\pi - \delta_\pm\right),
\end{equation}
which is indeed finite, and becomes
$\Delta\varphi=2\pi - \delta_+=2\pi - \delta_-$ in the $\alpha\to 1$
limit.
Thus, the geometry of extra dimensions is nothing but a round sphere
with a deficit angle $\delta_+=\delta_-$, i.e. a football-shaped
extra-dimensions considered in \cite{Carroll-Guica,Garriga-Porrati}.

As we shall see in the next section, when the new coordinates
$\left(\bar r,\varphi\right)$ are used and the KK mass is properly
rescaled, the system of the perturbation equations depends on the
background parameters through just one parameter $\alpha$. 
Thus, we calculate the mass spectra of KK modes in each type of
perturbations while changing $\alpha$ from 1 to 0. 
The perturbation equations written in terms of 
$\left(\bar r,\varphi\right)$ remain regular in the $\alpha\to 1$ limit.

\section{Dynamical stability}
\label{stability}

In this section, we consider linear perturbations around the background
spacetime described in the previous section in order to show the
dynamical stability of this spacetime.
For simplicity we assume that the perturbations are axisymmetric in the
extra dimensions. 
Since the background geometry has the 4-dimensional Poincare symmetry,
it is convenient to expand perturbations by scalar-, vector- and
tensor-type harmonics of the 4-dimensional Minkowski spacetime.
The perturbations are labeled by its type and values of mass squared
$m^2 \equiv - \eta^{\mu\nu} k_\mu k_\nu$, where $ k_\mu $ is the 4D projected
wave number of each harmonics.

To study the stability, we take advantages of the fact that for any
perturbation type, the perturbation equations are reduced to eigenvalue
problems with eigenvalue $m^2$. 
We regard the background spacetime to be dynamically stable if the
spectrum of $m^2$ is real and non-negative.
In each sector, our strategy to tackle this problem consists of the
following four steps:
\begin{itemize}
 \item [(i)] We show the reality of $m^2$. 
 \item [(ii)] We rewrite the systems of the perturbation equations into
       a form which depends on background parameters through just one
       parameter $\alpha$. The parameter $\alpha$ runs over the finite
       interval $\left[0,1\right]$. 
 \item [(iii)] In the $\alpha\to 1$ limit we analytically solve the
       perturbation equations and show that the spectrum of $m^2$ is
       non-negative. 
 \item [(iv)] We numerically evaluate how each eigenvalue $m^2$ changes
       as $\alpha$ changes from $1$ to $0$, and show that the spectrum
       remains non-negative throughout.
\end{itemize}
The steps (i) and (ii) make it possible for us to analyze the stability
in the $\alpha$-$m^2$ plane, i.e. a $2$-dimensional space spanned by
the paramter $\alpha$ and the eigenvalue $m^2$. In particular, each
eigenvalue generates a curve in the $\alpha$-$m^2$ plane as $\alpha$
runs over the interval $[0,1]$, and what we have to show is that all
such curves are in the stable region $m^2\geq 0$. Thus, provided  that
each eigenvalue is a continuous function of $\alpha$, the step (iii)
implies that, if there exists a value of $\alpha$ ($0\leq\alpha\leq 1$)
for which the background is unstable, the curve of the lowest eigenvalue
must cross the $\alpha$ axis at least once. Finally, the step (iv)
proves the stability of the system. In this way, we shall show that our 
six-dimensional brane world model is dynamically stable against tensor-,
vector- and scalar-type linear perturbations. While we can easily show
the stability against tensor-type perturbations even without solving the
perturbed Einstein equation, we need to perform numerical calculations
in the step (iv) for scalar- and vector-type perturbations. For
completeness, we shall perform numerical calculations for tensor
perturbations as well and show first few KK modes.

In the analysis of each sector, we shall require regularity of
physically relevant, geometrical quantities such as the Ricci scalar of
the induced metric on the brane, the tetrad components of the bulk Weyl
tensor evaluated on the brane, etc. This is because we are adopting the
thin brane approximation, i.e. (\ref{eqn:deficit-angle-formula}), and
all we can and should trust is what is obtained within the validity of
this approximation. If e.g. the Ricci scalar of the induced metric were
singular then our approximation would be invalidated. It is of course
possible to regularize the singularity by introducing a finite thickness
of the brane. However, in this case the natural cutoff of the low energy
effective theory is the inverse of the thickness, and in general the
regularized ``would-be singularity'' is not expected to be below the
cutoff scale. This simply means that we need a UV completion, e.g. the
microphysical description of the brane, to describe the physics of the
regularized ``would-be singularity''. Therefore, in general we have two
options: (a) to specify a fundamental theory such as string theory as a
UV completion and go on; or (b) to concentrate on modes which are
within the validity of the effective theory. In this paper we adopt the
latter attitude, assuming the existence of a good UV completion but
never using its properties. This is the reason why we adopt the thin
brane approximation and require the regularities.

Let us now start the analysis. To begin with, we expand the perturbed
metric by harmonics of the 4-dimensional Minkowski spacetime as
%
%
\begin{eqnarray}
&&\delta g_{MN} dx^M dx^N = h_{rr}Ydr^2 + 2h_{r\phi}Ydrd\phi
+ h_{\phi\phi}Yd\phi^2 
\nonumber\\&&~~~
+2\biggl( h_{(T)r} V_{(T)\mu} + h_{(L)r} V_{(L)\mu} \biggr)drdx^\mu
+2\biggl( h_{(T)\phi} V_{(T)\mu} + h_{(L)\phi} V_{(L)\mu} \biggr)
d\phi dx^\mu
\nonumber\\&&~~~
+\biggl( h_{(T)}T_{(T)\mu\nu} + h_{(LT)}T_{(LT)\mu\nu} +
h_{(LL)}T_{(LL)\mu\nu} + h_{(Y)}T_{(Y)\mu\nu} \biggr)dx^\mu dx^\nu,
\end{eqnarray}
where $Y$, $V_{(T,L)}$ and $T_{(T,LT,LL,Y)}$ are scalar, vector and
tensor harmonics, respectively.
%
%
%
%
%
See Appendix~\ref{harmonics} for definitions of the harmonics.
The coefficients $h_{rr}$, $h_{r\phi}$, $h_{\phi\phi}$, $h_{(T,L)r}$,
$h_{(T,L)\phi}$ and $h_{(T,LT,LL,Y)}$ are supposed to depend only on
$r$. In the same manner, the perturbations of the $U(1)$ gauge field can
be expanded as
%
%
\begin{eqnarray}
\delta A_M dx^M = a_r Y dr + a_\phi Y d\phi 
+ \biggl(a_{(T)}V_{(T)\mu} + a_{(L)}V_{(L)\mu}\biggr)dx^\mu.
\end{eqnarray}
Here the coefficient $a_r$, $a_\phi$ and $a_{(T,L)}$ are also supposed to
depend only on $r$ due to axisymmetry.

The Einstein equations and the Maxwell equations are decomposed into
three groups, each of which contains only variables belonging to one of
the following three sets of variables $\{h_{(T)}\}$, 
$\{h_{(T)r}, h_{(T)\phi}, h_{(LT)}, a_{(T)}\}$
and
$\{h_{rr},h_{r\phi},h_{\phi\phi},h_{(L)r},h_{(L)\phi},h_{(LL,Y)},a_r
,a_\phi,a_{(L)}\}$.
Variables belonging to each set are called tensor type, vector
type and scalar type, respectively.
It should be noted that these variables include degrees of freedom of
gauge transformation, which are explicitly given in
Appendix~\ref{gauge}.

\subsection{Tensor-type perturbation}
\label{tensor}

Our first task is to show that $m^2\geq 0$ for any non-vanishing
tensor perturbations.
We first derive the evolution equation for
tensor perturbations in subsection~\ref{basic_tensor}.
From this equation and regularity of $h$ and $h'$, we can show
$m^2\geq 0$ without solving the evolution equation.
We also show in subsection~\ref{basic_tensor} that the system of the
evolution equation and the boundary condition can be rewritten in the
form which depends on the background parameters through just one
parameter $\alpha \equiv r_-/r_+$. 
Then, we seek the analytic solution for $\alpha=1$ in
subsection~\ref{analytic_tensor}. Using this result, we numerically solve 
the perturbation equations by relaxation method in
subsection~\ref{kkmode}. Detailed explanation of relaxation method is given
in Appendix~\ref{relax}.

\subsubsection{Basic equations}
\label{basic_tensor}

For tensor perturbations, 
%
\begin{eqnarray}
 ds_6^2 & = & r^2 (\eta_{\mu\nu} + h Ys_{\mu\nu})
  dx^{\mu}dx^{\nu} + \frac{dr^2}{f}
  + fd\phi^2,\nonumber\\ 
 A_Mdx^M & = & Ad\phi,
\end{eqnarray}
where the perturbation is specified by the function $h$ of $r$
and the harmonics $Y\equiv\exp(ik_{\mu}x^{\mu})$. The symmetric
polarization tensor $s_{\mu\nu}$ satisfies
$s^{\mu}_{\mu}=k^{\mu}s_{\mu\nu}=0$ for $k^{\mu}k_{\mu}\ne 0$, or 
$s^{\mu}_{\mu}=k^{\mu}s_{\mu\nu}=\tau^{\mu}s_{\mu\nu}=0$ for
$k^{\mu}k_{\mu}=0$, where $\tau^{\mu}$ is a constant timelike
vector. There is no relevant equation coming from the Maxwell equation,
and the perturbed Einstein equation becomes 
%
\begin{equation}
 \frac{1}{r^2}\left(r^4 f h'\right)' + m^2 h = 0,
\label{tensor_eq}
\end{equation}
where a prime denotes derivative with respect to $r$ and
$m^2\equiv -\eta^{\mu\nu}k_{\mu}k_{\nu}$.
The regularity of the
four-dimensional Ricci scalar of the induced metric on the brane requires
that $h$ should be regular. With the above equation for $h$, the
regularity of the tetrad components of the six-dimensional Weyl tensor
on the brane requires that $h'$ should also be regular. Indeed, provided
that $h$ is regular, $C^r_{\ \mu r\nu}/r$ is regular at $r=r_{\pm}$ if
and only if $h'$ is regular.

With the regularity of $h$ and $h'$ at $r=r_{\pm}$, it is easy to show
that $m^2\geq 0$ for any non-vanishing solutions: 
%
\begin{equation}
 m^2\int_{r_-}^{r_+}dr\ r^2h^2  = 
  -\int_{r_-}^{r_+}dr\ h\left(r^4 f h'\right)' 
  = \int_{r_-}^{r_+}dr\ (r^2h')^2 f \geq 0,
\end{equation}
where we have done integration by parts and used the fact that
$f(r_{\pm})=0$. The equality holds if and only if $h'=0$ in the region
$r_-\leq r\leq r_+$. Thus, there is no instability in the tensor sector
and the zero mode is $h=const$.

The above differential equation (\ref{tensor_eq}) include the
two background parameters $(\mu_b,b)$, or equivalently $(r_+,r_-)$.
From now, we show this equation can be rewritten into a form
which includes only one parameter $\alpha=r_-/r_+$.
First of all, we perform a coordinate transformation Eq.(\ref{rbar}).
In terms of $\bar r$, the Eq.(\ref{tensor_eq}) becomes
%
\begin{eqnarray}
\frac{1}{\left(\beta_- \bar r +\beta_+\right)^2} \partial_{\bar r}
\left( \left(\beta_- \bar r +\beta_+\right)^4
\bar f  \partial_{\bar r}h \right) + \tilde m^2 h=0,
\label{tensor_eq2}
\end{eqnarray}
where we defined
%
\begin{eqnarray}
\tilde m^2 &=& \frac{m^2}{r_+ r_-}.
\end{eqnarray}
The function $\bar f$ and thus the Eq.(\ref{tensor_eq2})
include only the parameter $\alpha$.
The boundary conditions for $h$ are obtained by assuming that $h$ can be
expanded in the Taylor series at $r=r_\pm$:
%
\begin{equation}
\left. \partial_{\bar r}h + \frac{\tilde m^2}
{\left(\beta_- \bar r +\beta_+\right)^2 \partial_{\bar r} \bar f}h
\right|_{\bar r\to {\pm}1}=0.
\label{bc_tensor}
\end{equation}
Thus, it is sufficient that we solve the perturbation equation with a
variety of $\alpha$.

\subsubsection{Analytic solution for $\alpha = 1$}
\label{analytic_tensor}

Here, we give the solution of $h$ for $\alpha=1$, which is used when
we numerically solve the perturbation for general $\alpha$ in the next
subsection.
Substituting $\bar f$
into Eq.(\ref{tensor_eq2}) and taking $\alpha\to
1$, we get
%
\begin{eqnarray}
\partial_{\bar r} \left( \left(1-\bar r^2\right) 
\partial_{\bar r} h\right) + \mu^2 h=0,
\label{tensor_eq3}
\end{eqnarray}
where
%
\begin{eqnarray}
\mu^2 \equiv \frac{\tilde m^2}{2\Lambda_6}.
\end{eqnarray}
%
This can be solved analytically as
%
\begin{eqnarray}
h = C P_\nu \left(\bar r\right) + D Q_\nu \left(\bar r\right),
\label{sol_tensor}
\end{eqnarray}
where $C$ and $D$ are normalization constants.
$P_\nu$ and $Q_\nu$ are the Legendre functions of the first and second
kind, respectively.
$\nu$ is related to $\mu^2$ as $\mu^2 =\nu(\nu+1)$.
The regularity of $h$ at $\bar r=1$ requires $D=0$, and that at
$\bar r=-1$ does $P_\nu$ not to diverge there.
This is realized if only if $\nu$ is non-negative integer.
The mass spectrum is then determined as
%
\begin{eqnarray}
\mu^2 = \nu(\nu+1) ~~~~\left(\nu=0,1,2,\cdots\right).
\label{sp_tensor}
\end{eqnarray}
The zero mode (a mode with $m^2=0$) is $h=const.$ and all other modes
(i.e. Kaluza-Klein modes) have positive $m^2$. Thus, the background
spacetime for $\alpha =1$ is dynamically stable against tensor
perturbations.

\subsubsection{Numerical solution of KK modes}
\label{kkmode}

Here we obtain the first few KK modes of tensor-type perturbation by
numerically solving the perturbed Einstein equation (\ref{tensor_eq2})
with the boundary condition (\ref{bc_tensor}).
For this purpose, the relaxation method is useful.
Detailed explanation of the relaxation method is given in
Appendix~\ref{relax}.
Here we give an outline of this method.
We first rewrite the second order differential equation
(\ref{tensor_eq2}) to a system of two first order differential
equations by defining $\partial_{\bar r} h$ as well as $h$ as a
dependent variable.
Then, the differential equations are replaced by
finite-difference equations on a mesh of $M$ points that covers the
range of the integration.
We start with an arbitrary trial solution which does not necessarily 
satisfy the desired finite-difference equations, nor
the required boundary conditions.
The successive iteration, now called relaxation, will adjust all the
values on the mesh so as to realize a closer
agreement with finite-difference equations and, simultaneously, with the
boundary conditions.
Good initial guesses are the key of efficiency in the relaxation method.
Here we have to solve the problem many times, each time with a slightly
different value of $\alpha$.
In this case, the previous solution will be a good initial guess when
$\alpha$ is changed, and it will make relaxation work well.
As shown in the previous subsection, the perturbation equations can be
analytically solved for $\alpha=1$.
Thus, we solve the problem while changing $\alpha$ from 1 to 0.



\begin{figure}[tbp]
\leavevmode
\begin{center}
\includegraphics[width=7cm]{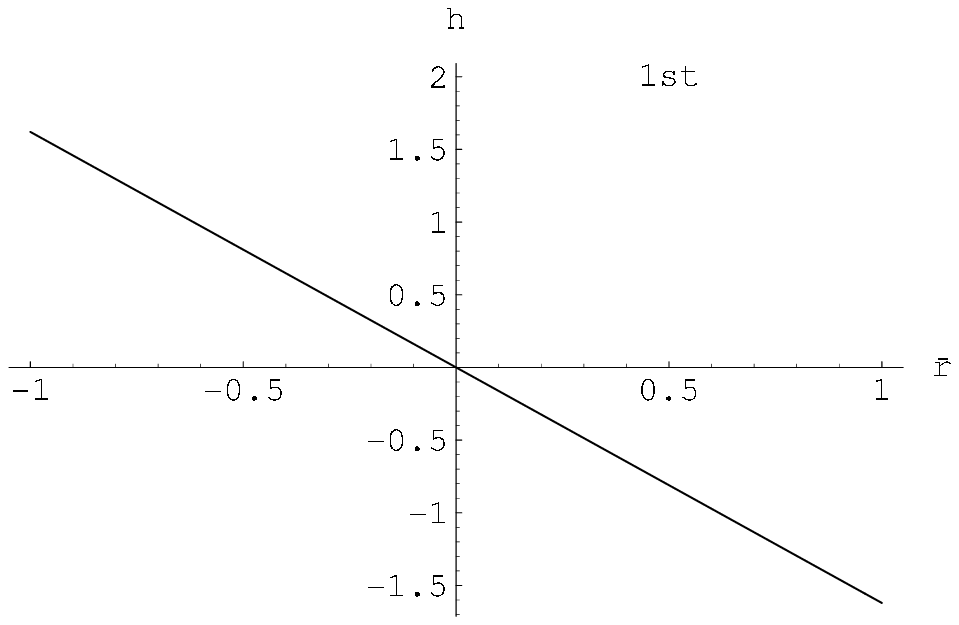}
\includegraphics[width=7cm]{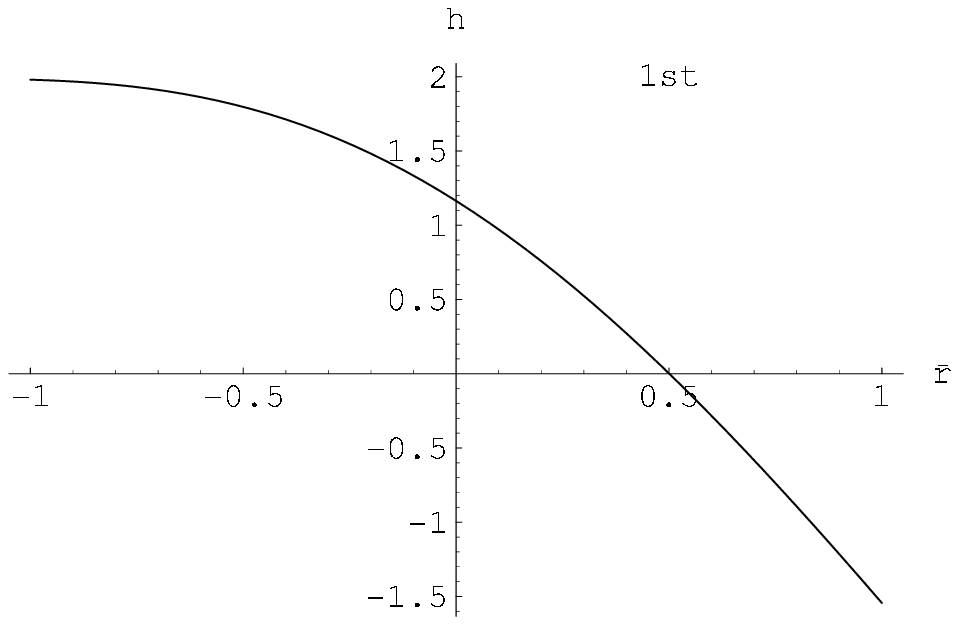}
\includegraphics[width=7cm]{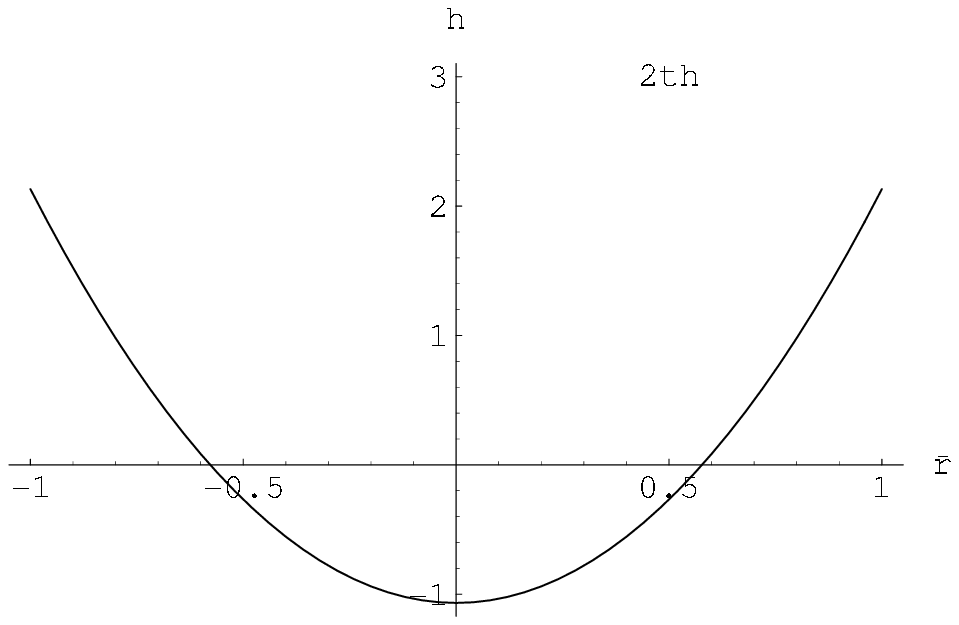}
\includegraphics[width=7cm]{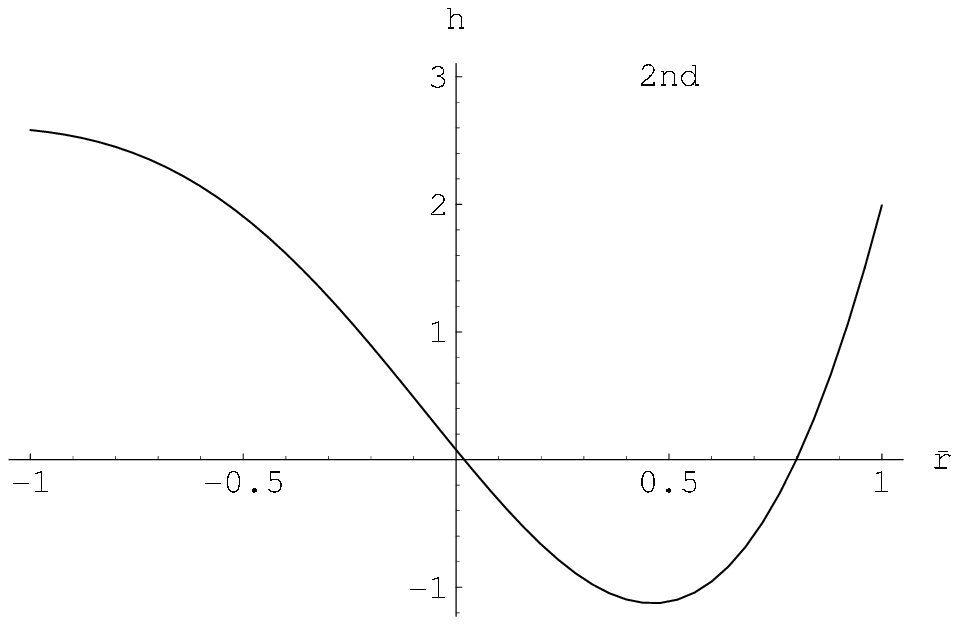}
\includegraphics[width=7cm]{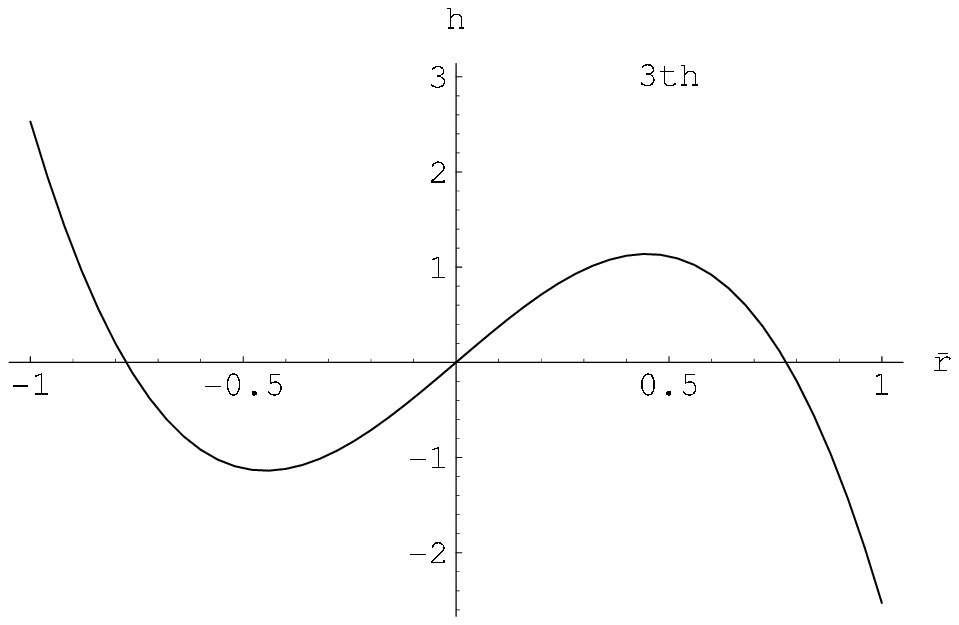}
\includegraphics[width=7cm]{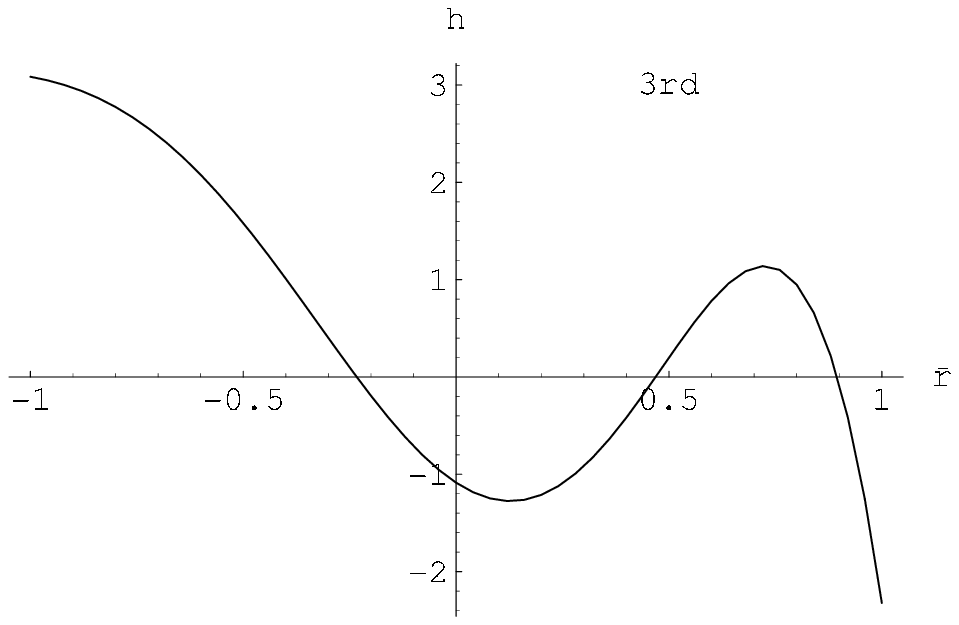}
\includegraphics[width=7cm]{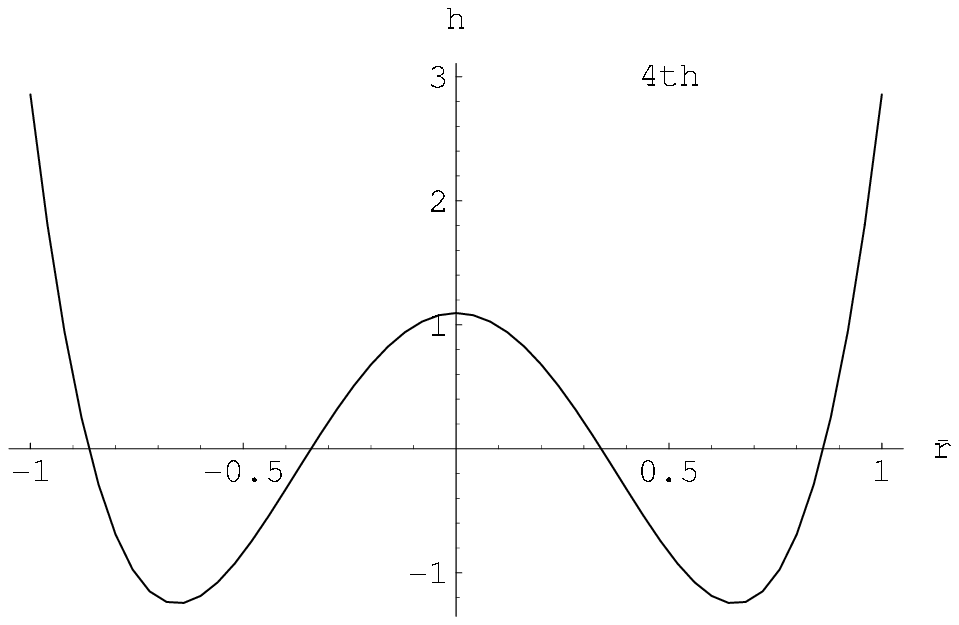}
\includegraphics[width=7cm]{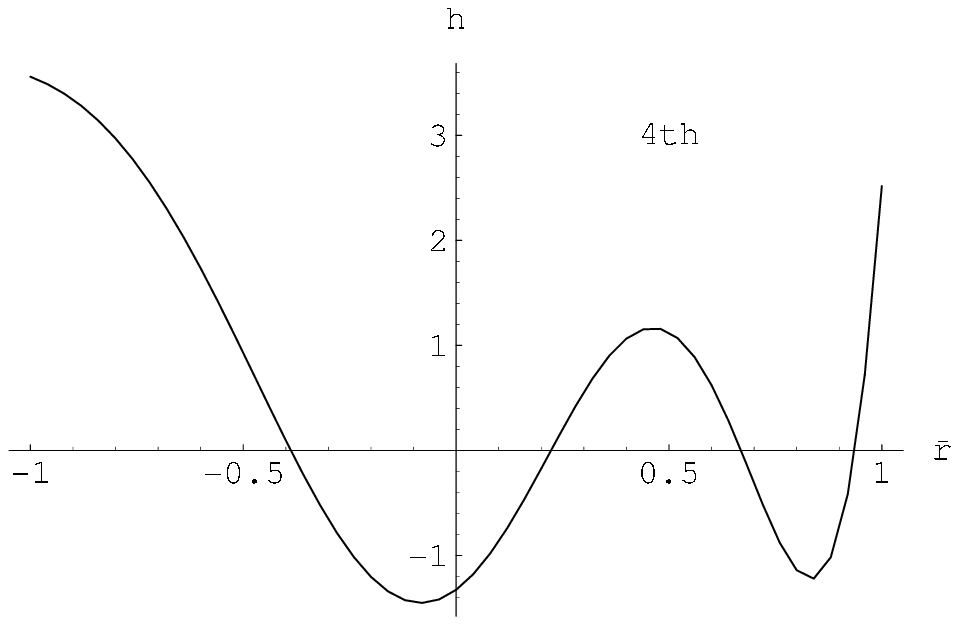}
\caption{The solution $h$ of the first four KK modes for $\alpha=1$
 (left) and $\alpha=0.11$ (right).
The normalization is determined by using the generalized Klein-Gordon
 norm (see the text).
Number of points of the mesh is taken to be $51$.
}
\label{fig:tensor_KK}
\end{center}
\end{figure}

We show the first four KK mode solutions of $h$ for a given value of
$\alpha$ in figure~\ref{fig:tensor_KK}.
When we plot the solutions, normalization is determined by using the
generalized Klein-Gordon norm.
See Appendix~\ref{KGnorm} for its derivation.
For tensor perturbations, it is defined by
%
\begin{equation}
 (\Phi,\Psi)_{KG} \equiv -i \frac{M_6^4\Delta\phi}{8}
   \int d^3{\bf x}\int dr r^2  \eta^{\mu\mu'}\eta^{\nu\nu'}
  \left(\Phi_{\mu\nu}\partial_t\Psi^*_{\mu'\nu'}
   -\Psi^*_{\mu\nu}\partial_t\Phi_{\mu'\nu'}\right).
\label{KG_tensor}
\end{equation}
Substituting
$\Phi_{\mu\nu}=\int d^4 k ~h_k s_{\mu\nu}Y$ and
$\Psi_{\mu'\nu'}=\int d^4 k' ~h_{k'} s_{\mu'\nu'}Y$, we get
%
\begin{eqnarray}
\left(\Phi,\Psi\right)_{KG} =
\left(k_0 + k'_0\right)\delta^3 \left({\bf k}-{\bf k'}\right)
\frac{M_6^4  \Delta\phi}{8}
 \int dr r^2 h_{n_1}(r)h_{n_2}(r),
\end{eqnarray}
where subscript $n_k$ means $h_{n_k}$ is the solution of tensor
perturbations with the eigenvalue $m_{n_k}^2$, and we normalized the
symmetric polarization tensor as $s_{\mu\nu}s^{\mu\nu}=1$.
In terms of the coordinate $\left(\bar r,\varphi\right)$, this becomes
%
\begin{eqnarray}
\left(\Phi,\Psi\right)_{KG} &=&
\left(k_0 + k'_0\right)\delta^3 \left({\bf k}-{\bf k'}\right)
\Delta\varphi \frac{M_6^4 r_+ r_-}{16\Lambda_6}
\int d\bar r \frac{\left(\beta_- \bar r + \beta_+\right)^2}{4}
h_{n_1}(\bar r)h_{n_2}(\bar r)
\nonumber\\
&\equiv & \left(k_0 + k'_0\right)\delta^3 \left({\bf k}-{\bf k'}\right)
\Delta\varphi \frac{M_6^4 r_+ r_-}{16\Lambda_6}
\left(h_{n_1},h_{n_2}\right).
\end{eqnarray}
We normalize the solution by
%
\begin{eqnarray}
\frac{\left(h_{n_1},h_{n_2}\right)}{\left(1,1\right)}
=\delta_{n_1 n_2},
\end{eqnarray}
so that the zero mode solution is normalized as $h(\bar r)=1$.
We can easily prove the orthogonality between modes with different $m^2$
by using the equation of motion for $h$, (\ref{tensor_eq2}).

\begin{figure}[t]
\centerline{\includegraphics[width=12cm]{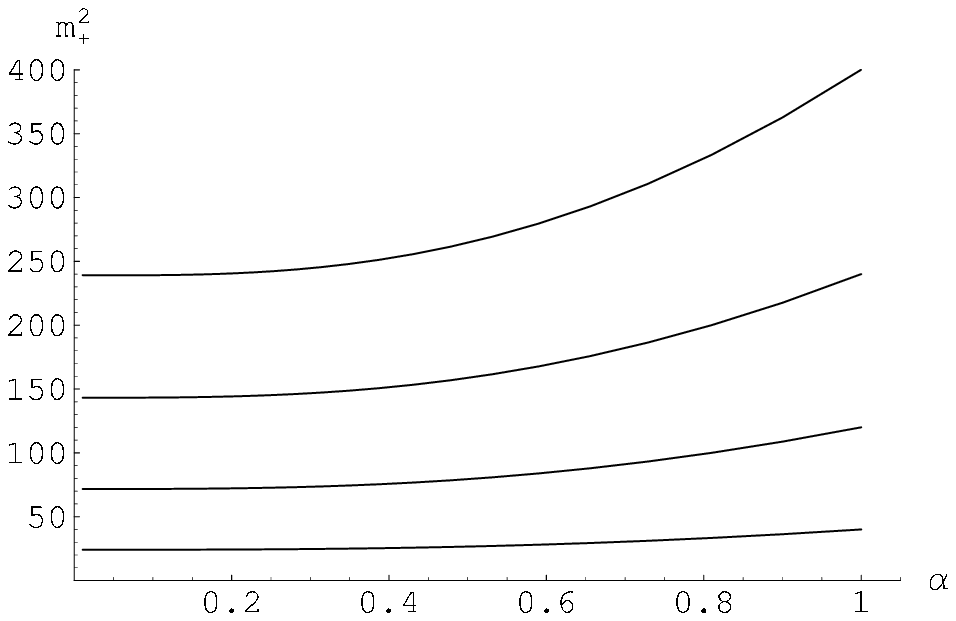}}
\caption{The spectrum of $m_+^2$ for tensor perturbations as a function
 of $\alpha$.}
\label{fig:sp_tensor}
\end{figure}

Finally, we show the spectrum of the mass squared of first four KK modes
as a function of $\alpha$.
The mass squared that has physical meaning is
$m_\pm^2\equiv -r_\pm^{-2}\eta^{\mu\nu}k_\mu k_\nu$,
which are the ones observed on the brane at $r=r_\pm$.
They are related to $\tilde m^2$ as
$m_\pm^2 = \alpha^{\pm 1}\tilde m^2$.
We plot the spectrum of $m_+^2$ in figure~\ref{fig:sp_tensor}.
As is easily seen, $m_+^2$ is non-negative for the entire range of
$\alpha$.
Therefore, the background spacetime we consider is dynamically stable in
the tensor-type sector.

We also notice that $m_+^2$ remains finite in $\alpha\to 0$ limit, which
implies $m_+^2\propto\alpha^0$, whereas $m_-^2\propto\alpha^{-2}$ for
$\alpha\to 0$.
This result is consistent with our previous study~\cite{paper1}, where
we analyzed how the Hubble expansion rate $H_{\pm}$ on each brane
changes when the tension of the brane changes.
In that paper
we also considered higher-order corrections of the effective Friedmann
equation with respect to $H_{\pm}$.
The result is that the higher-order corrections appear when $H_{\pm}$ get
larger than critical values $H_{*\pm}$.
For $\alpha\to 0$, we found that $H_{*\pm}$ behave as
%
\begin{eqnarray}
H_{*+}^2 \propto \alpha^0,~~~~H_{*-}^2 \propto \alpha^{-2}.
\label{Hbehave}
\end{eqnarray}
Since the higher-order corrections are caused by the KK modes,
this energy scale corresponds to their mass measured on each brane.
Thus, the behavior of $m_\pm^2$ we obtained in the $\alpha\to 0$ is
consistent with our previous result in~\cite{paper1}.

\subsection{Vector-type perturbation}\label{vector}

Next we show that $m^2 > 0$ for any non-vanishing
vector perturbations satisfying relevant boundary conditions.
We first derive the perturbed Einstein equations 
in subsection~\ref{basic_vector}.
There are two physical degrees of freedom for vector perturbations.
Their differential equations are easily written into a form which is
manifestly hermite,
therefore $m^2$ is real.
They are also written into a form including only the background
parameter $\alpha$.
We then derive the boundary conditions from those equations, by assuming
that the two perturbation variables can be Taylor expanded with respect
to $\bar r \pm 1$. In subsection~\ref{analytic_vector}, we summarize the
analytic solution for $\alpha=1$. Using this result, we numerically
solve the perturbation equations by relaxation method in
subsection~\ref{kkmode_vector}.

\subsubsection{Basic equations}\label{basic_vector}

For vector perturbations, 
%
\begin{eqnarray}
 ds_6^2 & = & r^2 \eta_{\mu\nu}dx^{\mu}dx^{\nu} 
  + 2(h_{(T)r}dr + h_{(T)\phi}d\phi)V_{(T)\mu}dx^{\mu}
+ \frac{dr^2}{f}
  + fd\phi^2,\nonumber\\ 
 A_Mdx^M & = & a_{(T)}V_{(T)\mu}dx^{\mu} +  Ad\phi,
\end{eqnarray}
where the perturbation is specified by the functions \{$h_{(T)r}$,
$h_{(T)\phi}$, $a_{(T)}$\} of $r$.
As for gauge fixing, see Appendix~\ref{gauge}.
The $(LT)$- and $(T)r$-components of
the Einstein equation give
%
\begin{equation}
 h_{(T)r} = \frac{C}{r^2f},
\end{equation}
where $C$ is an arbitrary constant for $m^2=0$, or $C=0$ for 
$m^2\ne 0$. 
The $(T)$-component of the Maxwell equation and the $(T)\phi$-component
of the Einstein equation are reduced to 
%
\begin{eqnarray}
 \left(r^2f \Phi_{(V)1}'\right)' - \sqrt{2}r^4A'\Phi_{(V)2}'
  +m^2\Phi_{(V)1} & = & 0, \nonumber\\
 \left(r^6 \Phi_{(V)2}'\right)' + \sqrt{2}r^4A'\Phi_{(V)1}'
  +\frac{m^2r^4}{f}\Phi_{(V)2} & = & 0, 
  \label{eqn:master-eq-vector}
\end{eqnarray}
where
%
\begin{eqnarray}
 \Phi_{(V)1} & \equiv & \sqrt{2} a_{(T)}, \nonumber\\
 \Phi_{(V)2} & \equiv & \frac{h_{(T)\phi}}{r^2}.
\end{eqnarray}

For $m^2=0$, if the constant $C$ were nonzero then by going to the gauge
$\bar{h}_{(T)r}=0$ by the gauge transformation (\ref{eqn:gauge-tr-hTr}),
the metric component $\bar{h}_{(LT)}$ given by (\ref{eqn:gauge-tr-hLT})
would diverge on the brane. Thus, the regularity of the induced metric
requires that $C=0$ for $m^2=0$. (For $m^2\ne 0$, it has already been
shown that $C=0$.)

The regularity of the field strength for the pull-back of $A_Mdx^M$ on
the brane requires that $\Phi_{(V)1}$ is regular at $r=r_{\pm}$. 
With the above equations (\ref{eqn:master-eq-vector}), the regularity of
the tetrad components of the six-dimensional Weyl tensor on the brane
requires that $\Phi_{(V)2}/f$, $\Phi_{(V)2}'$ and $\sqrt{f}\Phi_{(V)1}'$
should be finite at $r=r_{\pm}$. These regularity conditions are enough
to make equations (\ref{eqn:master-eq-vector}) hermite. Therefore, $m^2$ is
real.

As in the case of tensor perturbations, we need to rewrite the above
equations (\ref{eqn:master-eq-vector}) into a form which includes only
$\alpha$ and which makes it possible for us to take the $\alpha\to 1$
limit without any divergence. Actually, while the variables
$\Phi_{(V)1}$ and $\Phi_{(V)2}$ were useful to show the reality of
$m^2$, for numerical calculation there is a more convenient choice of 
variables. The $(T)\varphi$-component of the metric perturbation
$h_{(T)\varphi}$ in the coordinate $\left(\bar r,\varphi\right)$ is
related to $h_{(T)\phi}$ as
%
\begin{eqnarray}
h_{(T)\phi} = \Lambda_6 \sqrt{r_+ r_-}\beta_- h_{(T)\varphi}.
\end{eqnarray}
The coefficient $\beta_-$ in the above equation vanishes in the
$\alpha\to 1$ limit.
Thus, we rescale $h_{(T)\phi}$ as
%
\begin{eqnarray}
\tilde h_{(T)\phi} = \frac{1}{\sqrt{\Lambda_6 r_+ r_-}\beta_-}
 h_{(T)\phi},
\end{eqnarray}
which approaches to $\sqrt{\Lambda_6}h_{(T)\varphi}$ in the
$\alpha\to 1$ limit.
Using this variable, we can rewrite the above equations
(\ref{eqn:master-eq-vector}) as
%
\begin{eqnarray}
&&\partial_{\bar r}^2 a_{(T)}+
\left(\frac{\partial_{\bar r}\bar f}{\bar f}
+ \frac{2\beta_-}{\beta_- \bar r +\beta_+}\right)
\partial_{\bar r}a_{(T)}
+\frac{8\sqrt{2}\Lambda_6}
{\bar f \left(\beta_- \bar r +\beta_+\right)^4}
\sqrt{\frac{3}{\gamma_1}}\sqrt{\frac{\gamma_2}{5}}
\biggl(\partial_{\bar r}\tilde h_{(T)\phi}
\nonumber\\&&~~~~~~~~~~~~
-\frac{2\beta_-}{\beta_- \bar r +\beta_+}\tilde h_{(T)\phi}\biggr)
+\frac{\tilde m^2}{\bar f \left(\beta_- \bar r +\beta_+\right)^2}
a_{(T)}=0
\nonumber\\&&
\partial_{\bar r}^2 \tilde h_{(T)\phi}
+ \frac{2\beta_-}{\beta_- \bar r +\beta_+}\partial_{\bar r}
 \tilde h_{(T)\phi}
-\frac{16\sqrt{2}}{\left(\beta_- \bar r +\beta_+\right)^4}
\sqrt{\frac{3}{\gamma_1}}\sqrt{\frac{\gamma_2}{5}}
\partial_{\bar r}a_{(T)}
\nonumber\\&&~~~~~~~~~~~~
-6\left(\frac{\beta_-}{\beta_- \bar r +\beta_+}\right)^2
\tilde h_{(T)\phi}
+\frac{\tilde m^2}{\bar f \left(\beta_- \bar r +\beta_+\right)^2}
\tilde h_{(T)\phi}=0,
  \label{eqn:master-eq-vector2}
\end{eqnarray}
where $\gamma_n =\sum_{i=0}^{2n}\alpha^{i-n}$ as we defined above.
The boundary conditions are obtained by assuming that $\Phi_{(V)1}$ and
$\tilde\Phi_{(V)2}$ can be expanded in the Taylor series at $r=r_\pm$:
%
\begin{eqnarray}
\left. \tilde h_{(T)\phi}
\right|_{\bar r\to {\pm}1}=0
\nonumber\\
\partial_{\bar r}a_{(T)}
+\frac{8\sqrt{2}\Lambda_6}
{\left(\beta_- \bar r +\beta_+\right)^4 \partial_{\bar r}\bar f}
\sqrt{\frac{3}{\gamma_1}}\sqrt{\frac{\gamma_2}{5}}
\biggl(\partial_{\bar r}\tilde h_{(T)\phi}
-\frac{2\beta_-}{\beta_- \bar r +\beta_+}\tilde h_{(T)\phi}\biggr)
\nonumber\\
+\frac{\tilde m^2}
{\left(\beta_- \bar r +\beta_+\right)^2\partial_{\bar r}\bar f}
a_{(T)}
\biggr|_{\bar r\to {\pm}1}=0.
\label{bc_vector}
\end{eqnarray}

\subsubsection{Analytic solution for $\alpha = 1$}
\label{analytic_vector}

The system of the perturbation equations (\ref{eqn:master-eq-vector2})
becomes simple in the $\alpha\to 1$ limit, and then can be solved
analytically.
Here we summarize those solutions, which are used when we numerically
solve the equations (\ref{eqn:master-eq-vector2}) and (\ref{bc_vector})
for general $\alpha$.
Taking $\alpha\to 1$ and using the equation (\ref{fbar_limit}), the
equations (\ref{eqn:master-eq-vector2}) become
%
\begin{eqnarray}
 \partial_{\bar r}\left[(1-\bar r^2)\partial_{\bar r}a_{(T)}\right]
+\sqrt{2\Lambda_6}\partial_{\bar r}h_{(T)\varphi} + \mu^2 a_{(T)} & = & 0,
  \nonumber\\
 \partial_{\bar r}^2 h_{(T)\varphi}
- \sqrt{\frac{2}{\Lambda_6}}\partial_{\bar r} a_{(T)}
+ \frac{\mu^2}{1-\bar r^2}h_{(T)\varphi} & = & 0,
\end{eqnarray}
where
%
\begin{eqnarray}
\mu^2  \equiv  \frac{\tilde m^2}{2\Lambda_6}. 
\end{eqnarray}
The absence of the zero mode solution can be easily shown using the
above differential equations and the regularity of the variables at the
boundaries~\cite{Sendouda}.
In the following, we consider the case of $\mu^2 \ne 0$.

By the change of variables
%
\begin{eqnarray}
 \Psi_1 & \equiv & a_{(T)}, \nonumber\\
 \Psi_2 & \equiv & -\sqrt{2\Lambda_6}\partial_{\bar r}h_{(T)\varphi}
+ 2a_{(T)},
\end{eqnarray}
the above equations become
%
\begin{eqnarray}
 \partial_{\bar r}\left[(1-\bar r^2)\partial_{\bar r}\Psi_1\right]
 -(\Psi_2-2\Psi_1) +\mu^2\Psi_1 & = & 0, 
  \nonumber\\
 \partial_{\bar r}\left[(1-\bar r^2)\partial_{\bar r}\Psi_2\right]
 + \mu^2(\Psi_2-2\Psi_1) & = & 0. 
\end{eqnarray}
The original variables are written in terms of $\Psi_1$ and $\Psi_2$ as
%
\begin{eqnarray}
a_{(T)} & = & \Psi_1
= \frac{1}{2\mu^2}
\left\{\partial_{\bar r}\left[(1-\bar r^2)\partial_{\bar r}\Psi_2\right]
+ \mu^2\Psi_2\right\}, \nonumber\\
h_{(T)\varphi} & = & \frac{1-\bar r^2}{\sqrt{2\Lambda_6}\mu^2}
\partial_{\bar r}\Psi_2.
\end{eqnarray}
This set of equations can be rewritten as
%
\begin{equation}
\partial_{\bar r} \left[(1-\bar r^2)\partial_{\bar r}E_{\pm}\right]
 + \lambda_{\pm}E_{\pm} = 0,
\end{equation}
where
%
\begin{eqnarray}
 E_{\pm} & \equiv & 
\partial_{\bar r}\left[(1-\bar r^2)\partial_{\bar r}\Psi_2\right]
 + \lambda_{\mp}\Psi_2,
  \nonumber\\
 \lambda_{\pm} & \equiv & \mu^2+1\pm\sqrt{2\mu^2+1}.
\end{eqnarray}
Thus, the general solution is
%
\begin{equation}
 \Psi_2 = C_+P_{\nu_+}(\bar r) + C_-P_{\nu_-}(\bar r)
  + D_+Q_{\nu_+}(\bar r) + D_-Q_{\nu_-}(\bar r),
\end{equation}
where $C_{\pm}$ and $D_{\pm}$ are constants and $\nu_{\pm}$ is a
solution to 
%
\begin{equation}
 \nu_{\pm}(\nu_{\pm}+1) = \lambda_{\pm}. 
\end{equation}

From the regularity of $\Phi_{(V)1}$, $\sqrt{f}\Phi_{(V)1}'$,
$\Phi_{(V)2}/f$ and $\Phi_{(V)2}'$, we can show that $\Psi_1$,
$\sqrt{1-\bar r^2}\partial_{\bar r}\Psi_1$, $\Psi_2$ and
$\partial_{\bar r}\Psi_2$ should be finite at the boundaries.
At $\bar r=1$, $\Psi_2$ is regular only if $D_+ + D_- =0$ whereas the
regularity of $\Psi_1$ is reduced through the differential equations to
$\left(\lambda_+ -\mu^2\right)D_+ +\left(\lambda_- -\mu^2\right)D_- =0$.
Since $\lambda_+ \ne \lambda_-$ for $\mu^2 \ne -1/2$, we obtain
$D_+ = D_- =0$.
At $\bar r=-1$, the regularity conditions for the variables become
%
\begin{eqnarray}
&& C_+ \sin\nu_+\pi + C_- \sin\nu_-\pi = 0, \nonumber\\
&& C_+ \lambda_+ \sin\nu_+\pi + C_- \lambda_- \sin\nu_-\pi = 0.
\end{eqnarray}
Non-trivial solutions can exist if $ \nu_+ \in \mathbb Z $ or
$ \nu_- \in \mathbb Z $.
In general we can choose non-negative $ \nu_\pm $'s so that these
conditions are explicitly written down as
%
\begin{equation}
\nu_\pm
 = \frac{-1+\sqrt{4\mu^2+5 \pm 4\sqrt{2\mu^2+1}}}{2} = 0,1,2,\ldots.
\end{equation}
Therefore the KK mass spectrum for vector perturbation is obtained.
The case of $\mu^2 \ne -1/2$ can be excluded~\cite{Sendouda}.

\subsubsection{Numerical solution of KK modes}
\label{kkmode_vector}

Here we obtain the first few KK modes of vector type perturbations by
numerically solving the perturbed Einstein equation
(\ref{eqn:master-eq-vector2}) and the junction condition
(\ref{bc_vector}).
We rewrite the system of two second order differential equations
(\ref{eqn:master-eq-vector2}) to a system of four first order
differential equations by defining $\partial_{\bar r} a_{(T)}$ and
$\partial_{\bar r} \tilde h_{(T)\phi}$ as well as $a_{(T)}$ and
$\tilde h_{(T)\phi}$ as dependent variables.
Here, we take the number of points on a mesh $M=101$.
We solve the problem while changing $\alpha$ from 1 to 0, each time with
a slightly different value of $\alpha$.
Starging from $\alpha=1$, the analytic solutions for $\alpha=1$
presented in the previous subsection are used as a trial solution.

\begin{figure}[tbp]
\leavevmode
\begin{center}
\includegraphics[width=7cm]{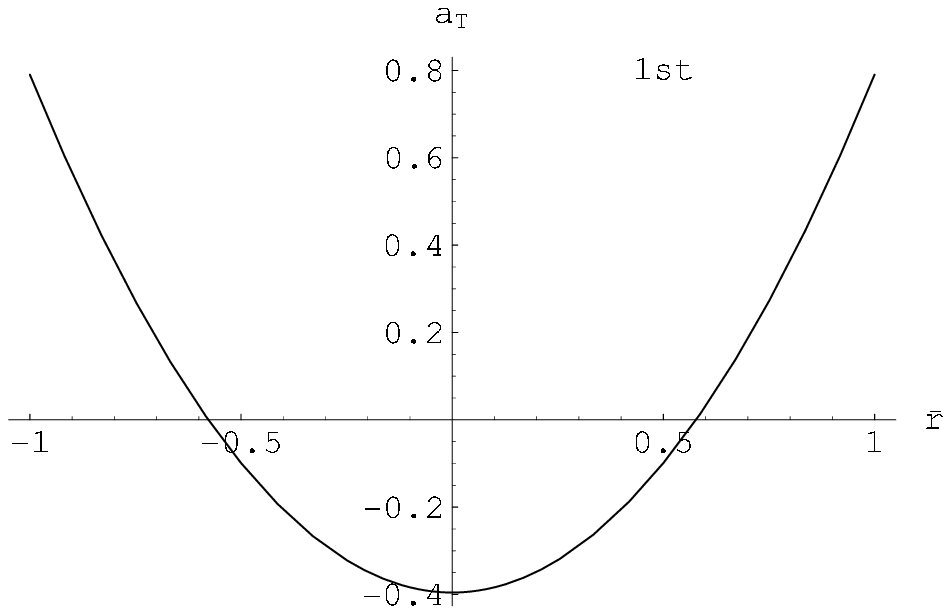}
\includegraphics[width=7cm]{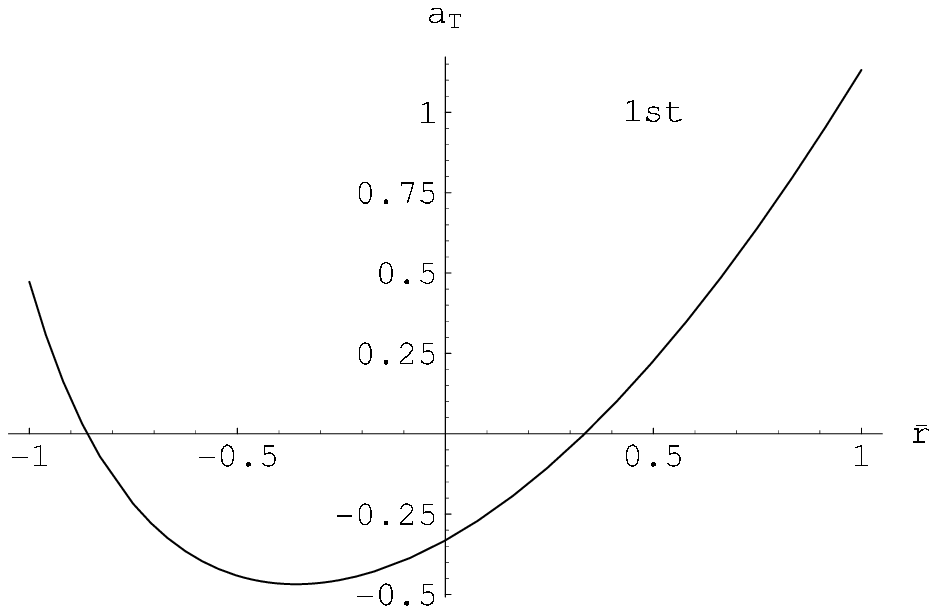}
\includegraphics[width=7cm]{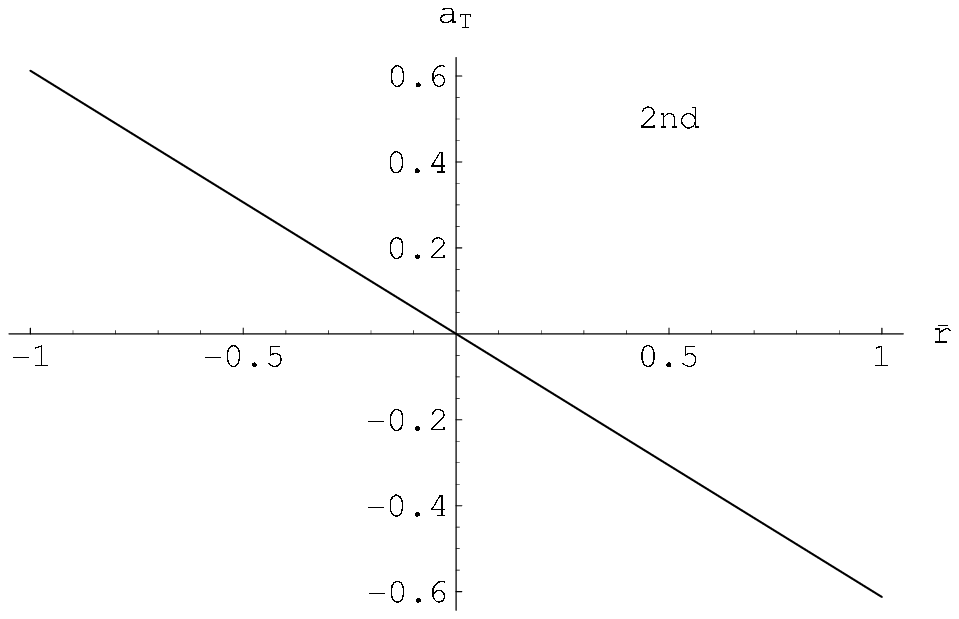}
\includegraphics[width=7cm]{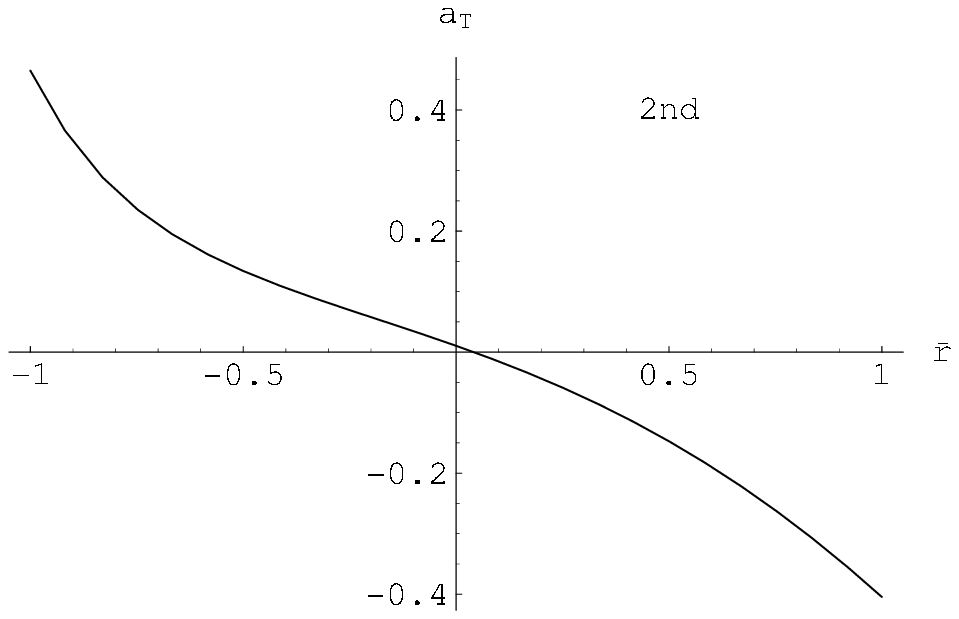}
\includegraphics[width=7cm]{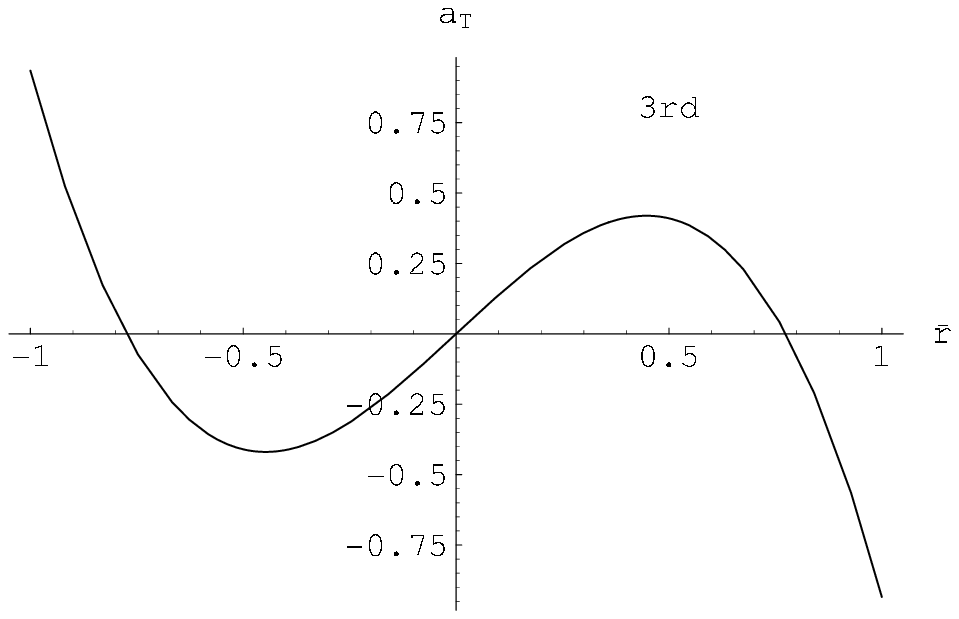}
\includegraphics[width=7cm]{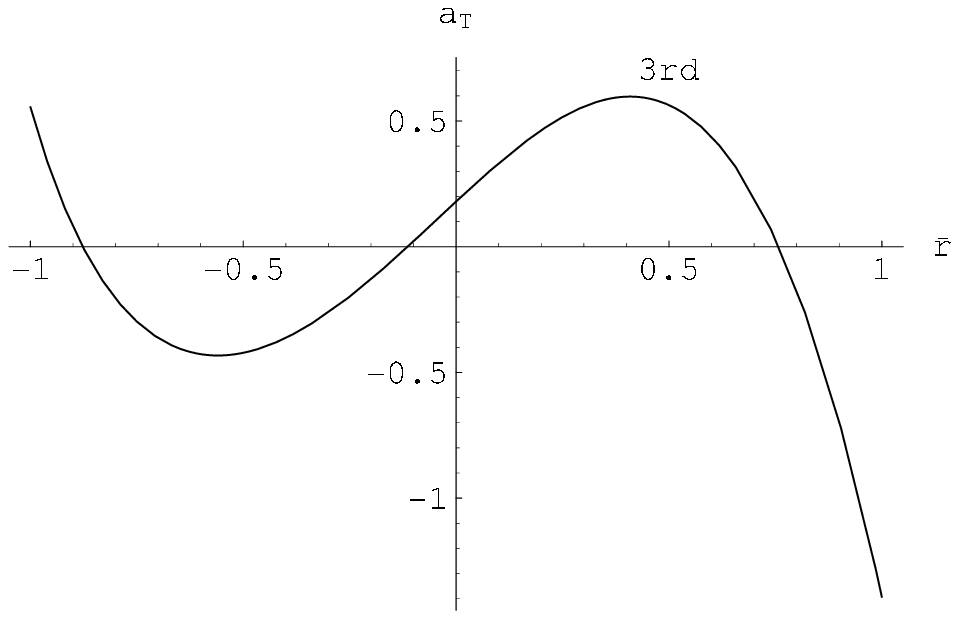}
\includegraphics[width=7cm]{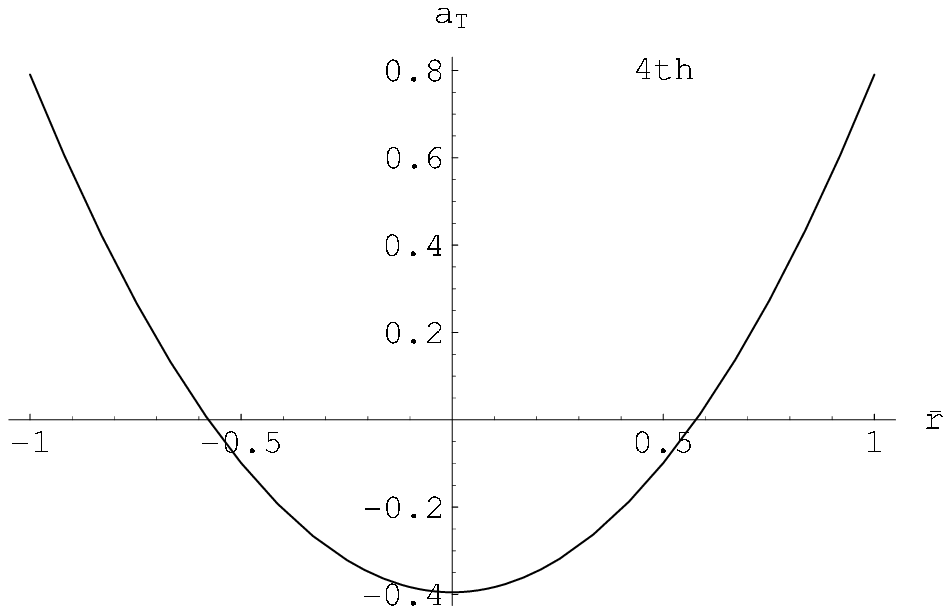}
\includegraphics[width=7cm]{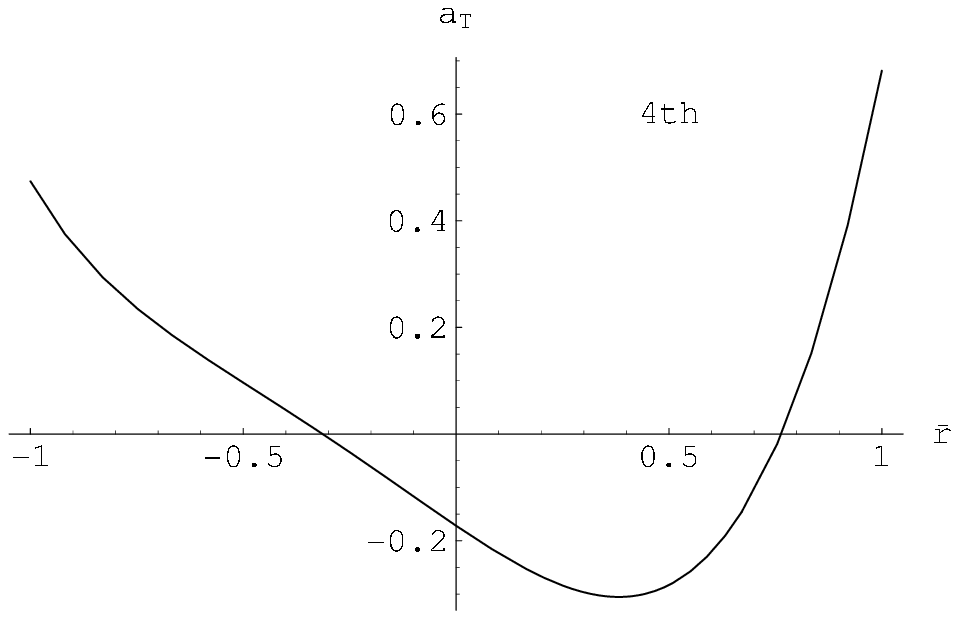}
\caption{The solution $a_{(T)}$ of the first four KK modes for $\alpha=1$
 (left) and $\alpha=0.31$ (right).
The normalization is determined by using the generalized Klein-Gordon
 norm (see the text).
Number of points of the mesh is taken to be $101$.
}
\label{fig:vector_a}
\end{center}
\end{figure}

\begin{figure}[tbp]
\leavevmode
\begin{center}
\includegraphics[width=7cm]{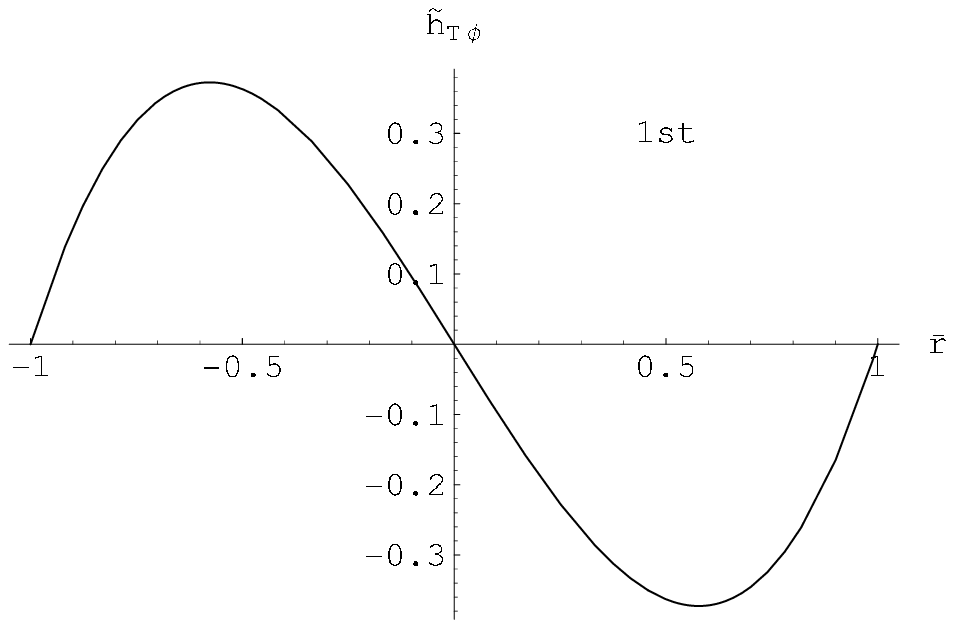}
\includegraphics[width=7cm]{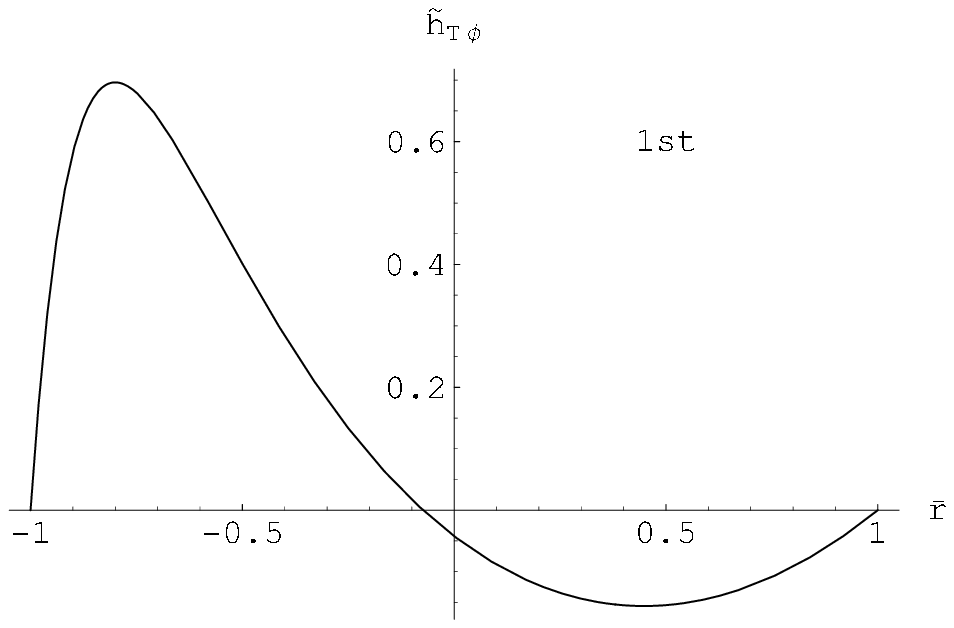}
\includegraphics[width=7cm]{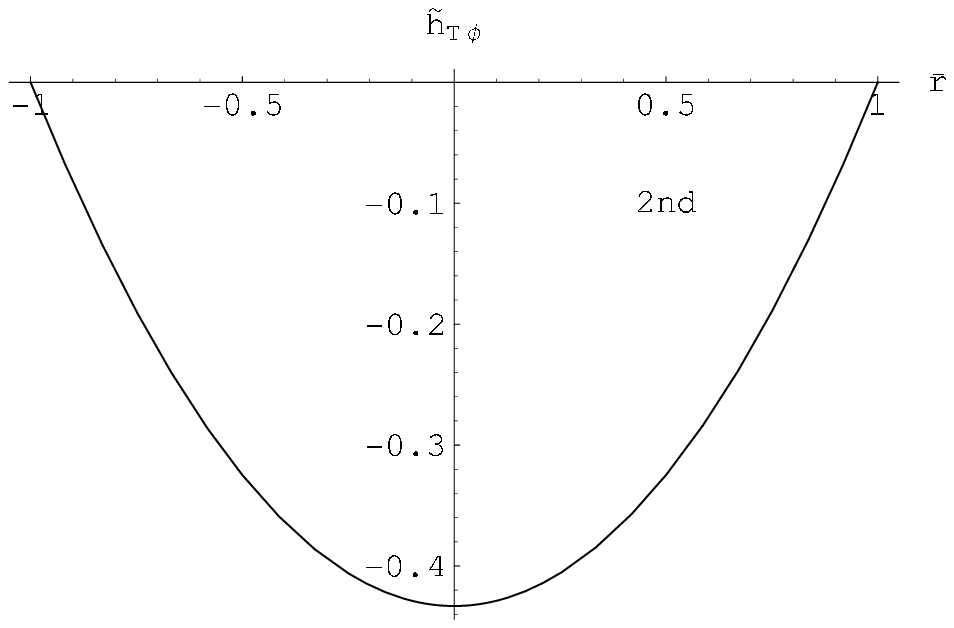}
\includegraphics[width=7cm]{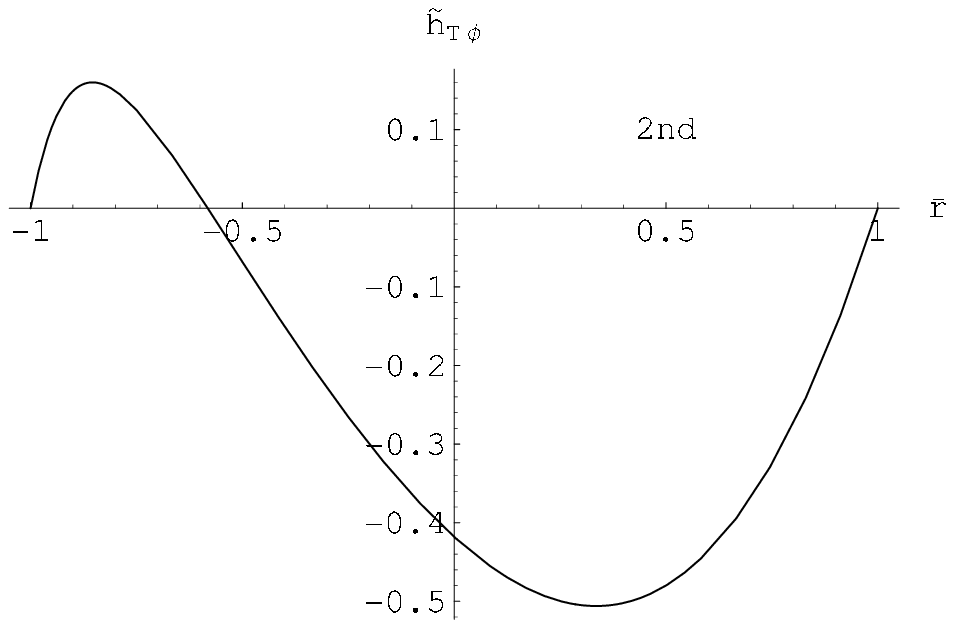}
\includegraphics[width=7cm]{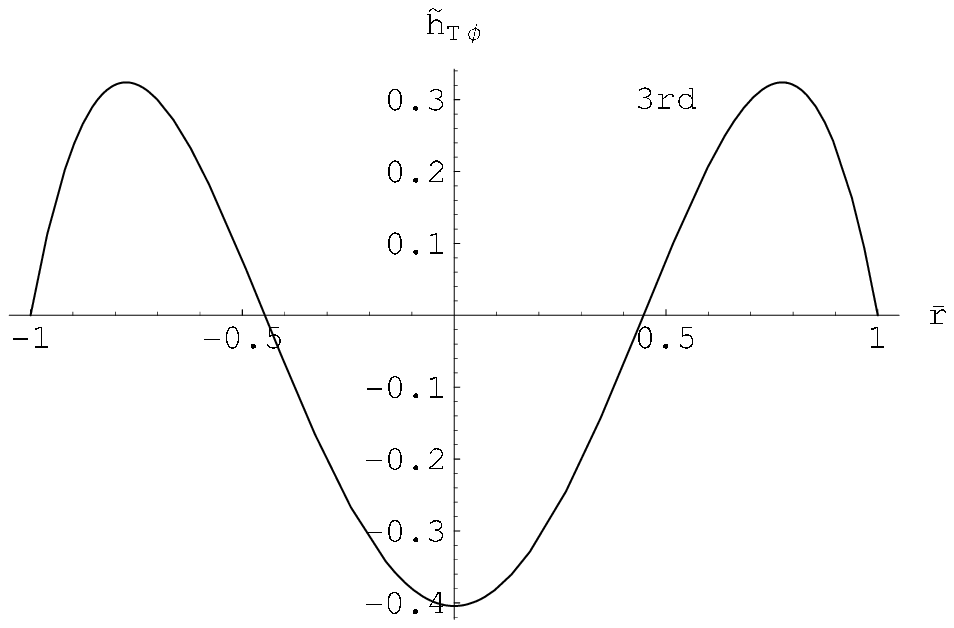}
\includegraphics[width=7cm]{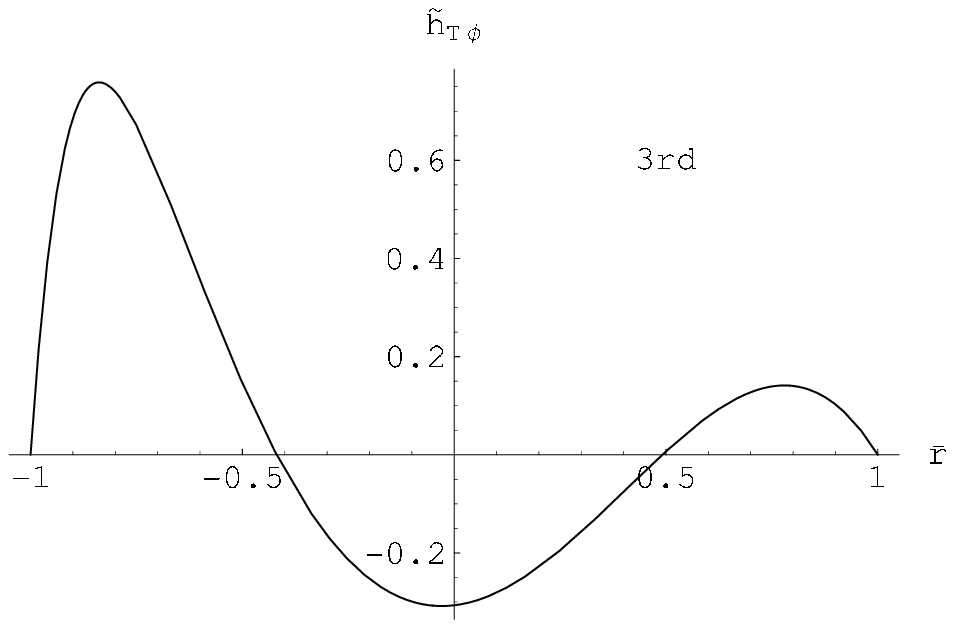}
\includegraphics[width=7cm]{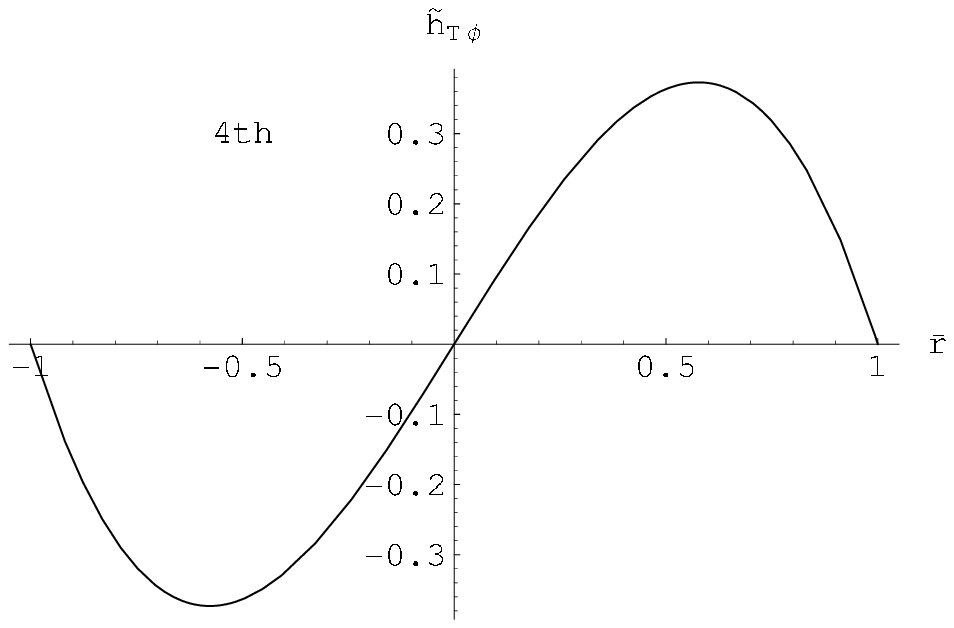}
\includegraphics[width=7cm]{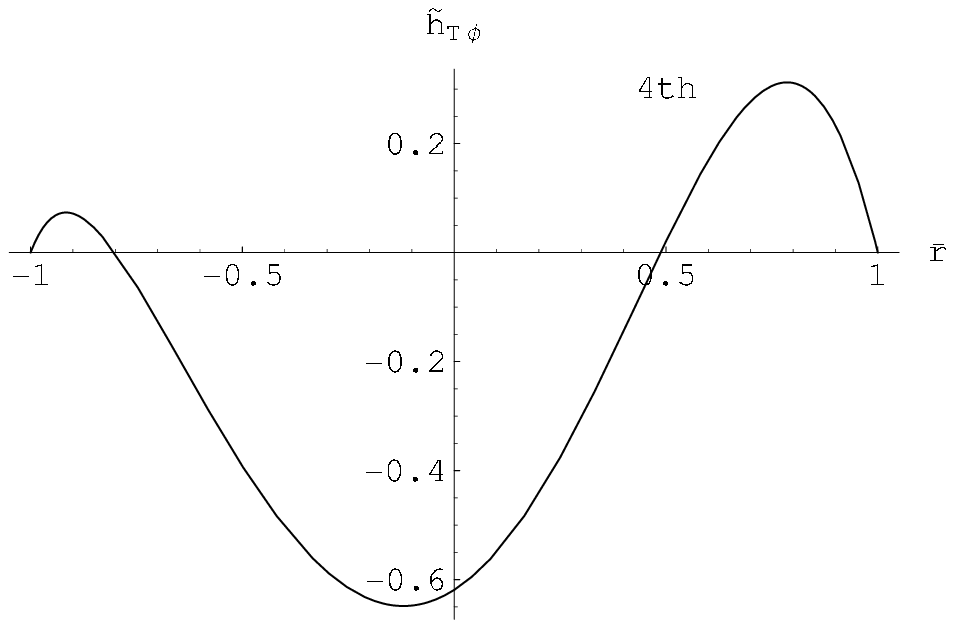}
\caption{The solution $\tilde h_{(T)\phi}$ of the first four KK modes
 for $\alpha=1$ (left) and $\alpha=0.31$ (right).
The normalization is determined by using the generalized Klein-Gordon
 norm (see the text).
Number of points of the mesh is taken to be $101$.
}
\label{fig:vector_h}
\end{center}
\end{figure}

Figure~\ref{fig:vector_a} and \ref{fig:vector_h} show the first four KK
mode solutions of $a_{(T)}$ and $\tilde h_{(T)\phi}$ for $\alpha=1.0$
and $0.31$.
The normalization is determined by using the generalized Klein-Gordon
norm as in tensor perturbations.
For vector perturbations, it is defined by
%
\begin{equation}
 (\tilde{\Phi},\tilde{\Psi})_{KG} \equiv -i \frac{M_6^4\Delta\phi}{2}
  \int d^3{\bf x}\int dr \eta^{\mu\nu}
  \left[\left(\tilde{\Phi}_{1\mu}\partial_t\tilde{\Psi}_{1\nu}^*
	 -\tilde{\Psi}^*_{1\mu}\partial_t\tilde{\Phi}_{1\nu}\right)
   +\frac{r^4}{f}\left(\tilde{\Phi}_{2\mu}\partial_t\tilde{\Psi}_{2\nu}^* 
   -\tilde{\Psi}^*_{2\mu}\partial_t\tilde{\Phi}_{2\nu}\right)\right].
\end{equation}
See the Appendix~\ref{KGnorm} for the derivation.
In terms of $a_{(T)}$ and $h_{(T)\phi}$, this is written as
%
\begin{equation}
 (\tilde{\Phi},\tilde{\Psi})_{KG} =
\left(k_0 + k'_0\right)\delta^3\left({\bf k}-{\bf k'}\right)
\frac{M_6^4 \Delta\phi}{2}
\int dr \left[ 2a_{(T)}^{(n_1)}a_{(T)}^{(n_2)}
 + \frac{1}{f}h_{(T)\phi}^{(n_1)}h_{(T)\phi}^{(n_2)}
 \right],
\end{equation}
where superscript $n_k$ means that $a_{(T)}^{(n_k)}$ and 
$h_{(T)\phi}^{(n_k)}$ are the solutions of vector perturbations
with eigenvalue $m_{n_k}^2$, and we normalized the constant vector as
$u_\mu u^\mu=1$.
Using the coordinate $\left(\bar r,\varphi\right)$ and the rescaled
variable $\tilde h_{(T)\phi}^{(n_k)}$, the Klein-Gordon
norm is further rewritten as
%
\begin{eqnarray}
 (\tilde{\Phi},\tilde{\Psi})_{KG} &=&
\left(k_0 + k'_0\right)\delta^3 \left({\bf k}-{\bf k'}\right)
\frac{M_6^4\Delta\varphi}{4\Lambda_6}
\int d\bar r \left[ 2a_{(T)}^{(n_1)}a_{(T)}^{(n_2)}
 + \frac{\Lambda_6}{\bar f}\tilde h_{(T)\phi}^{(n_1)}
\tilde h_{(T)\phi}^{(n_2)} \right]
\nonumber\\
&\equiv & \left(k_0 + k'_0\right)\delta^3 \left({\bf k}-{\bf k'}\right)
\frac{M_6^4\Delta\varphi}{4\Lambda_6}
\left(a_{(T)}^{(n_1)},\tilde h_{(T)\phi}^{(n_1)}
\Big| a_{(T)}^{(n_2)},\tilde h_{(T)\phi}^{(n_2)}\right).
\end{eqnarray}
We normalize the solution by
%
\begin{eqnarray}
\left(a_{(T)}^{(n_1)},\tilde h_{(T)\phi}^{(n_1)}
\Big| a_{(T)}^{(n_2)},\tilde h_{(T)\phi}^{(n_2)}\right)
=\delta_{n_1 n_2}.
\end{eqnarray}
We can easily prove the orthogonality between modes with different $m^2$
by using the equation of motion for vector perturbations.

\begin{figure}[t]
\centerline{\includegraphics[width=12cm]{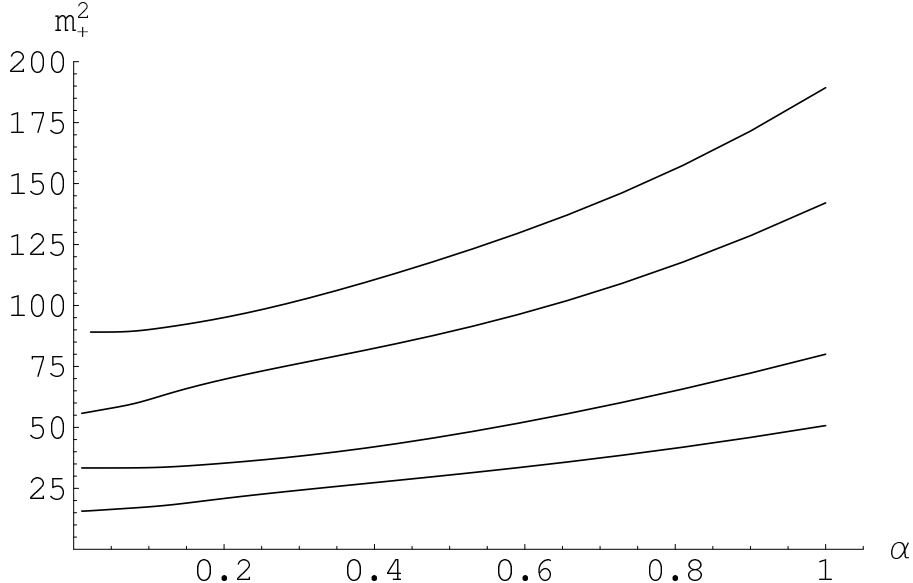}}
\caption{The spectrum of $m_+^2$ for vector perturbations as a function
 of $\alpha$.}
\label{fig:sp_vector}
\end{figure}

Finally, we show the spectrum of $m_+^2$ for the first four KK modes
as a function of $\alpha$ in figure~\ref{fig:sp_vector}.
We find that $m_+^2$ is non-negative for the entire range of
$\alpha$.
Therefore, the background spacetime is also dynamically
stable in the vector-type sector.

The behavior of $m_+^2$ in $\alpha\to 0$ limit has the same feature as
in tensor perturbations.
It remains finite in this limit, that is, $m_+^2\propto\alpha^0$.
As was mentioned in subsection~\ref{kkmode}, 
this result is consistent with our previous study~\cite{paper1}, where
we have found that the energy scale $H_{*+}$ at which the KK modes begin
to modify the effective Friedmann equation on the brane at $r=r_+$
behaves as $H_{*+}^2 \propto \alpha^0$ when $\alpha\sim 0$.

\subsection{Scalar-type perturbation}
\label{scalar}

Finally, we show that $m^2 > 0$ for any non-vanishing
scalar perturbations satisfying relevant boundary conditions.
We first derive the perturbed Einstein equations and the boundary
conditions in subsection~\ref{basic}.
As well as the vector perturbations, there are two physical degrees of freedom.
We then rewrite the system of the perturbation equations into a form
including only the background parameter $\alpha$.
In subsection~\ref{reality}, we show that $m^2$ is real.
We summarize the analytic solution for $\alpha=1$ in
subsection~\ref{analytic_scalar}.
Using this result, we numerically solve the perturbation equations for
$ \alpha<1 $ by relaxation method in subsection~\ref{kkmode_scalar}.

\subsubsection{Basic equations}
\label{basic}

For scalar perturbations, we can take an analog of the longitudinal gauge:
%
\begin{eqnarray}
 ds_6^2 & = & r^2(1+\Psi Y)
  \eta_{\mu\nu}dx^{\mu}dx^{\nu}
  + \left[1+(\Phi_1+\Phi_2)Y\right]\frac{dr^2}{f}
  + 2 h_{(L)\phi}\partial_{\mu}Ydx^{\mu}d\phi
\nonumber\\&&~~~~~~
  + \left[1-(\Phi_1+3\Phi_2)Y\right]fd\phi^2,\nonumber\\  
 A_Mdx^M & = & a_rYdr + (A+a_{\phi}Y)d\phi,
\end{eqnarray}
where perturbations are specified by the functions \{$\Psi$, $\Phi_1$,
$\Phi_2$, $h_{(L)\phi}$, $a_r$, $a_{\phi}$\} of $r$ and the harmonics 
$Y\equiv\exp(ik_{\mu}x^{\mu})$. As for gauge fixing, see
Appendix~\ref{gauge}. The $r\phi$-component of the Einstein equation
implies that $h_{(L)\phi}=Cf(r)$, where $C$ is an arbitrary constant. By
using the residual gauge freedom $\tilde{C}$ in
(\ref{residual-gauge-freedom}), we can set $C=0$. Thus, 
%
\begin{equation}
 h_{(L)\phi} = 0.
\end{equation}
The $r$-component of the Maxwell equation and the $(LL)$-, $(L)r$- and
$(L)\phi$-components of the Einstein equations give 
%
\begin{eqnarray}
 a_r & = & 0, \nonumber\\
 \Psi & = & \Phi_2, \nonumber\\
 A'a_{\phi} & = & \frac{1}{2r^2}(fr^2\Phi_1)'+ f'\Phi_2, 
\end{eqnarray}
where a prime denotes derivative with respect to $r$. The remaining
equations are reduced to  
%
\begin{eqnarray}
 \Phi_1'' + 2\left(\frac{f'}{f}+\frac{5}{r}\right)\Phi_1'
  - \frac{4\Lambda_6}{f}(\Phi_1+\Phi_2)
  + \frac{m^2}{r^2f}\Phi_1 & = & 0, \nonumber\\
 \Phi_2'' + \frac{4}{r}\Phi_2' 
  + \frac{m^2}{2r^2f}
  \left(\Phi_1+2\Phi_2\right) & = & 0, 
  \label{eqn:master-eq-scalar}
\end{eqnarray}
where $m^2\equiv -\eta^{\mu\nu}k_{\mu}k_{\nu}$.

With these equations, it is straightforward to show that linear
perturbations of $R$, $R^{MN}_{M'N'}R^{M'N'}_{MN}$, $R^{;M}R_{;M}$ and
$R^{KLMN;M'}R_{KLMN;M'}$ are independent linear combinations of
$f\Phi_1$, $(f\Phi_1)'$, $\Phi_2$ and $f'\Phi_2'-(m^2/2r^2)\Phi_1$ and
that the matrix made of the coefficients remains regular and invertible
in the $r\to r_{\pm}$ limit. Thus, $f\Phi_1$, $(f\Phi_1)'$, $\Phi_2$ and
$f'\Phi_2'-(m^2/2r^2)\Phi_1$ must be regular on the boundaries.

The correct boundary conditions can be obtained either by using the
formalism developed in Sendouda et al.~\cite{Sendouda} or by setting the
coefficients of $1/f$ in (\ref{eqn:master-eq-scalar}) to zero on the 
boundary, where $f$ vanishes. In the latter method, we obtain 
%
\begin{eqnarray}
 \left.
  2f'\Phi_1' - 4\Lambda_6(\Phi_1+\Phi_2) + \frac{m^2}{r^2}\Phi_1
 \right|_{r\to r_{\pm}}  & = & 0, \nonumber\\
 \left.
  \frac{m^2}{2r^2}(\Phi_1+2\Phi_2)
 \right|_{r\to r_{\pm}}  & = & 0. 
 \label{eqn:easy-bc}
\end{eqnarray}
By using the boundary conditions, it is shown that $\Phi_1$ and $\Phi_2$
have regular Taylor expansion w.r.t. $r-r_{\pm}$.
In practice, it is useful to note that for given
values of $\Phi_1(r_{\pm})$ and $\Phi_2'(r_{\pm})$,

\begin{eqnarray}
 \Phi_2(r_{\pm}) & = & -\frac{1}{2}\Phi_1(r_{\pm}), \nonumber\\
  \Phi_1'(r_{\pm}) & = & 
   \left(\Lambda_6-\frac{m^2}{2r_{\pm}^2}\right)
   \frac{\Phi_1(r_{\pm})}{f'(r_{\pm})}, \nonumber\\
 \Phi_2''(r_{\pm}) & = & -\frac{4}{r_{\pm}}\Phi_2'(r_{\pm})
   + \frac{m^2}{2r_{\pm}^2f'(r_{\pm})}
   \left[
   \left(\Lambda_6-\frac{m^2}{2r_{\pm}^2}\right)
   \frac{\Phi_1(r_{\pm})}{f'(r_{\pm})}
   +2\Phi_2'(r_{\pm})\right].
\label{scalar_Taylor}
\end{eqnarray}

The above derivation of the boundary condition and the Taylor expansion
is simple and easy to follow. An alternative and more rigorous
derivation is also possible by using the formalism developed in Sendouda
et al.~\cite{Sendouda}. Following the formalism, the straightforward 
calculation gives 
%
\begin{eqnarray}
 \left.
  \left(\sqrt{f}f'\right)'\delta r - \frac{1}{2}f^{3/2}f'h_{rr}
  + \sqrt{f}h_{\phi\phi}'\right|_{r\to r_{\pm}} & = & 0, \nonumber\\
 \left.
  \left[\sqrt{f}\left(\sqrt{f}f'\right)'\right]'\delta r
  - f^{3/2}\left(\sqrt{f}f'\right)'h_{rr}
 + \sqrt{f} (\sqrt{f} h_{\phi\phi}')'
  -\frac{1}{2}ff'\left(f h_{rr}\right)'\right|_{r\to r_{\pm}} 
 & = & 0, \nonumber\\
  \left. A'\delta r + a_{\phi} \right|_{r\to r_{\pm}} & = & 0,
\end{eqnarray}
where $h_{rr}=(\Phi_1+\Phi_2)/f$, $h_{\phi\phi}=-(\Phi_1+3\Phi_2)f$ and
$\delta r$ is defined by 
%
\begin{equation}
 f'\delta r + h_{\phi\phi} = 0.
\end{equation}
By using the regularity of $f\Phi_1$, $(f\Phi_1)'$, $\Phi_2$ and
$f'\Phi_2'-(m^2/2r^2)\Phi_1$ at $r=r_{\pm}$, the boundary conditions are
reduced to 
%
\begin{equation}
  \left. f\Phi_1\right|_{r=r_{\pm}} = 0, \quad 
  \left. 2f'\Phi_2 + (f\Phi_1)'\right|_{r=r_{\pm}} = 0.
  \label{bc_r} 
\end{equation}
This is actually equivalent to (\ref{eqn:easy-bc}). Indeed, the same
Taylor expansion, i.e. (\ref{scalar_Taylor}), follows also from
(\ref{bc_r}).

Heretofore, the system of the above differential equations and the
boundary conditions can be rewritten into a form which includes only the
parameter $\alpha$.
In terms of $\bar r$, the Eq.(\ref{eqn:master-eq-scalar}) becomes
%
\begin{eqnarray}
&&\partial_{\bar r}^2\Phi_1
+ 2\left(\frac{\partial_{\bar r}\bar f}{\bar f} + 
     5\frac{\beta_-}{\beta_- \bar r +\beta_+}
  \right)\partial_{\bar r}\Phi_1
  - \frac{\Lambda_6}{\bar f}(\Phi_1+\Phi_2)
  + \frac{\tilde m^2}{\left( \beta_- \bar r +\beta_+ \right)^2\bar f}
\Phi_1 =  0, \nonumber\\
&&\partial_{\bar r}^2\Phi_2 + 4\frac{\beta_-}{\beta_- \bar r +\beta_+}
\partial_{\bar r}\Phi_2
  + \frac{\tilde m^2}{2\left( \beta_- \bar r +\beta_+\right)^2\bar f}
  \left(\Phi_1+2\Phi_2\right)  = 0.
  \label{eqn:master-eq-scalar2}
\end{eqnarray}
The function $\bar f$ and thus the Eq.(\ref{eqn:master-eq-scalar2})
include only the parameter $\alpha$.
For numerical calculations in subsection~\ref{kkmode_scalar}, we use the
first two equations in (\ref{scalar_Taylor}) as boundary conditions for
$\Phi_1$ and $\Phi_2$. These boundary conditions are also rewritten in
terms of $\bar{r}$, $\tilde{m}$ and $\alpha$:
%
\begin{equation}
 \left.\Phi_1 + 2\Phi_2\right|_{\bar r=\pm 1} = 0, \quad 
  \left. \partial_{\bar r}\Phi_1 =
\left(\Lambda_6 - \frac{\tilde m^2}{2}\alpha^{\pm 1}\right)
\frac{1}{4\partial_{\bar r} \bar f}\Phi_1\right|_{\bar r=\pm 1}.
\label{bc_rbar}
\end{equation}
Thus, to show the dynamical stability of scalar type perturbations, we
calculate the spectrum of $\tilde m^2$ as a function of $\alpha$ by
using (\ref{eqn:master-eq-scalar2}) with (\ref{bc_rbar}), and show that
$\tilde m^2$ is non-negative throughout.

As stated in the beginning of this subsection, we need to follow the
four steps (i)-(iv) to show the stability. The first step (i) has not
yet been considered at all, while the set of differential equations 
(\ref{eqn:master-eq-scalar2}) with (\ref{bc_rbar}) is simple enough and 
expected to be useful for the remaining steps (ii)-(iv). 
Here we note that the Eq.(\ref{eqn:master-eq-scalar}) is not a manifestly
self-adjoint system and, thus, we do not yet know whether the eigenvalue
$\tilde m^2$ is real or not. In the next subsection~\ref{reality} we shall
reduce the system of differential equations to a manifestly self-adjoint
one and show that $\tilde m^2$ is indeed real. After that, we shall come
back to the equations (\ref{eqn:master-eq-scalar2}) with (\ref{bc_rbar})
again and perform numerical calculations.

\subsubsection{Reality of $m^2$ for scalar perturbation}
\label{reality}

To show the reality of $m^2$ for scalar perturbations, we suppose that
$m^2\ne 0 $ and adopt the following gauge: 
%
\begin{eqnarray}
 ds_6^2 & = & r^2(1+Q_1 Y)
  \eta_{\mu\nu}dx^{\mu}dx^{\nu}
  + 2 {\cal B}V_{(L)\mu}dx^{\mu}dr 
  + (1+{\cal A}Y)\frac{dr^2}{f}
  + (1-3Q_1Y)fd\phi^2,\nonumber\\ 
 A_Mdx^M & = & {\cal C}Ydr + (A+Q_2Y)d\phi,
\end{eqnarray}
where $Q_1$, $Q_1$, ${\cal A}$, ${\cal B}$ and ${\cal C}$ are functions
of $r$. In this gauge the branes are at 
$r=r_{\pm}+3ff'Q_1Y|_{r=r_{\pm}}$. Note that the corresponding metric in
the five-dimensional Einstein frame after reducing the $\phi$ direction
is  
%
\begin{equation}
 ds_{5(E)}^2  = r^2f^{1/3}\eta_{\mu\nu}dx^{\mu}dx^{\nu}
  + 2 f^{1/3}{\cal B}V_{(L)\mu}dx^{\mu}dr 
  + [1+({\cal A}-Q_1)Y]\frac{dr^2}{f^{2/3}}.
\end{equation}
Thus, in this gauge a constant-$r$ hypersurface in the
five-dimensional Einstein frame is flat. Note also that the
($\phi\phi$)-component of the six-dimensional metric and the
$\phi$-component of the $U(1)$ field behave as scalar fields in
five-dimension after reducing the $\phi$ direction. Therefore, $Q_1$ and
$Q_2$ represent perturbations of the scalar fields on the flat
hypersurface in the five-dimensional Einstein frame.

In the analysis of cosmological perturbations in the four-dimensional
Einstein theory, the so called Mukhanov variables play important
roles. The Mukhanov variables represent perturbations of the
scalar fields on flat hypersurfaces, and thus, are analogous to our
variables $Q_1$ and $Q_2$. Therefore, it is expected that the above
gauge choice simplifies the analysis of perturbations.

The $r$-component of the Maxwell equation and the $(LL)$- and
$(L)r$-components of the Einstein equations give the following algebraic
equations. 
%
\begin{eqnarray}
 {\cal C} & = & 0, \nonumber\\
 \frac{m^2}{r^2}{\cal B} & = & 
  \frac{2rf'}{6f+rf'}Q_1' - \frac{2rA'}{6f+rf'}Q_2' 
  - \frac{12(\Lambda_6 r^2+9f)(2f+rf')+r^2{f'}^2}{(6f+rf')^2}\frac{Q_1}{r}
  + \frac{8\Lambda_6r^3A'}{(6f+rf')^2}\frac{Q_2}{r},\nonumber\\
 {\cal A} & = & 
  \frac{3(2f-rf')}{6f+rf'}Q_1 + \frac{4rA'}{6f+rf'}Q_2, 
\end{eqnarray}
where a prime denotes derivative with respect to $r$ and
$m^2=-\eta^{\mu\nu}k_{\mu}k_{\nu}$. The remaining equations are reduced
to 
%
\begin{equation}
 \mbox{\boldmath $L Q$} = m^2\mbox{\boldmath $\Omega Q$},
\end{equation}
where
%
\begin{equation}
 \mbox{\boldmath $Q$} =  
  \left(\begin{array}{c}
	Q_1\\ Q_2 \end{array} \right),
  \end{equation}
and
%
\begin{eqnarray}
 \mbox{\boldmath $L$} & = & 
 \partial_r \mbox{\boldmath $\alpha$}\partial_r
 + \partial_r\mbox{\boldmath $\beta$}
 + \mbox{\boldmath $\beta$}\partial_r 
 + \mbox{\boldmath $\gamma$}, \nonumber\\
 \mbox{\boldmath $\alpha$} & = & 
  r^4\left(\begin{array}{cc}
   3f & 0 \\
	 0 & 1
	\end{array}
  \right), \nonumber\\
 \mbox{\boldmath $\beta$} & = & 
  \frac{3}{2}r^4A'\left(\begin{array}{cc}
   0 & -1 \\
	 1 & 0
	\end{array}
  \right), \nonumber\\
 \mbox{\boldmath $\gamma$} & = & 
  -\frac{16r^6\Lambda_6{A'}^2}{(6f+rf')^2}
  \left(\begin{array}{cc}
   a & b \\
	 b & 1
	\end{array}
  \right), \nonumber\\
 \mbox{\boldmath $\Omega$} & = & 
  \frac{r^2}{f}\left(\begin{array}{cc}
   3f & 0 \\
	 0 & 1
	\end{array}
	  \right),\nonumber\\
 a & = & 
  -\frac{9f(2f+f')}{r^2{A'}^2}
  + \frac{9(2\Lambda_6+{A'}^2)}{32\Lambda_6r^2{A'}^2},
  \nonumber\\
 b & = & \frac{3(2f-rf')}{4rA'}.
\label{eqn:def-LOmega}
\end{eqnarray}

With these equations, the linear perturbations of $R$, 
$R^{MN}_{M'N'}R^{M'N'}_{MN}$, $R^{;M}R_{;M}$ and 
$R^{KL}_{MN;M'}R_{KL}^{MN;M'}$ at the positions of the branes are 
independent linear combinations of $Q_1$, $Q_2$, $Q_1'$ and $Q_2'$ 
evaluated at $r=r_{\pm}$~\footnote{Note that the positions of the branes
are $r=r_{\pm}+3f'fQ_1Y|_{r=r_{\pm}}$ and we do not know a priori
whether $fQ_1$ vanishes at $r=r_{\pm}$ or not. Thus, for example, the
linear perturbation of $R$ at the positions of the branes is 
$\delta R+3{R^{(0)}}'f'fQ_1Y$ evaluated at $r=r_{\pm}$, where $R^{(0)}$ 
and $\delta R$ are the background value and the linear perturbation of 
the Ricci scalar, respectively. Similar statements hold also for the linear 
perturbations of $R^{MN}_{M'N'}R^{M'N'}_{MN}$, $R^{;M}R_{;M}$ and 
$R^{KLMN;M'}R_{KLMN;M'}$ at the positions of the branes. 
}. 
The matrix made of the coefficients is regular and invertible. Thus, 
$Q_1$, $Q_2$, $Q_1'$ and $Q_2'$ must be regular at the positions of
branes. With this regularity condition, it is shown by using the
formalism developed in ref.~\cite{Sendouda} that the boundary condition at
$r=r_{\pm}$ is 
%
\begin{equation}
 \left.fQ_1\right|_{r=r_{\pm}} = \left.Q_2\right|_{r=r_{\pm}} = 0.
  \end{equation}
Alternatively, the same boundary condition can be obtained from the
boundary condition (\ref{eqn:easy-bc}) and the relation
(\ref{eqn:relation-Q-Phi}) below. 
It is easy to show that the operator {\boldmath $L$} with this boundary
condition is hermite and that $m^2$ is real unless $Q_1=Q_2=0$
everywhere in the interval $r_-\leq r\leq r_+$.

The Mukhanov-type variables ($Q_1$, $Q_2$) can be written in terms of
the metric variables ($\Phi_1$, $\Phi_2$) in the analog of the
longitudinal gauge. We have shown that $m^2$ is real unless the former
variables vanish everywhere in the interval $r_-\leq r\leq r_+$. On the
other hand, we have used the latter variables in the numerical
calculations. The relations between the sets of variables are
%
\begin{eqnarray}
 Q_1 & = & \Phi_2 + \frac{2f\Phi_1}{6f+rf'}, \nonumber\\
 Q_2 & = & a_{\phi}
  + \frac{frA'\Phi_1}{6f+rf'},
  \label{eqn:relation-Q-Phi}
\end{eqnarray}
where it is understood that $a_{\phi}$ is expressed in terms of $\Phi_1$
and $\Phi_2$. Equivalently, ($\Phi_1$, $\Phi_2$) are written in terms of
($Q_1$, $Q_2$) as 
%
\begin{eqnarray}
 \Phi_1 & = & (6f+rf')\frac{{\cal B}}{r}, \nonumber\\
 \Phi_2 & = & Q_1 - \frac{2f{\cal B}}{r},
\end{eqnarray}
where it is understood that ${\cal B}$ is expressed in terms of $Q_1$
and $Q_2$. Therefore, if $Q_1=Q_2=0$ everywhere in the interval 
$r_-\leq r\leq r_+$ then $\Phi_1=\Phi_2=0$ everywhere in the interval
$r_-\leq r\leq r_+$. Therefore, we have shown that $m^2$ is real unless
$\Phi_1=\Phi_2=0$ everywhere in the interval $r_-\leq r\leq r_+$.

\subsubsection{Analytic solution for $\alpha = 1$}
\label{analytic_scalar}

The perturbation equations given in the subsection~\ref{basic} can be
analytically solved for $\alpha=1$ as shown in Sendouda et
al.\cite{Sendouda}.
Here we summarize the solution obtained there.
By taking the $\alpha\to1$ limit of the equations
(\ref{eqn:master-eq-scalar2}) and (\ref{bc_rbar})
we get
%
\begin{eqnarray}
 \left(1-\bar r^2\right)\partial_{\bar r}^2\Phi_1
 - 4\bar r \partial_{\bar r}\Phi_1
  - 2(\Phi_1+\Phi_2)
  + \mu^2\Phi_1 & = & 0, \nonumber\\
 \left(1-\bar r^2\right)\partial_{\bar r}^2\Phi_2
+ \frac{\mu^2}{2}\left(\Phi_1+2\Phi_2\right) & = & 0,
  \label{eqn:master-eq-scalar3}
\end{eqnarray}
and
%
\begin{equation}
 \left.\Phi_1 + 2\Phi_2\right|_{\bar r=\pm 1} = 0, \quad 
  \left. \left(1-\bar r^2\right)
\partial_{\bar r}\Phi_1\right|_{\bar r=\pm 1} = 0,
\label{bc_rbar2}
\end{equation}
where $ \mu^2 \equiv \tilde m^2/2\Lambda_6 $.
The solution for the zero mode can be excluded using the first condition
of (\ref{bc_rbar2}) and the regularity of the variables at the
boundaries~\cite{Sendouda}.
In the following, we consider the case of $\mu^2 \ne 0$.

The differential equations are combined to
%
\begin{equation}
\partial_{\bar r}[(1-\bar r^2) \partial_{\bar r}F_\pm]
 + \lambda_\mp F_\pm = 0,
\end{equation}
where
%
\begin{eqnarray}
F_\pm
 & = &\partial_{\bar r}[(1-\bar r^2) \partial_{\bar r}\Phi_2]
 + \lambda_\pm \Phi_2, \\
\lambda_\pm
 & = &\mu^2+1 \pm \sqrt{3\mu^2+1}.
\end{eqnarray}
If $ \mu^2 \neq -1/3 $, the solution is obtained as
%
\begin{eqnarray}
\Phi_2
 & = &C_+ P_{\nu_+} + C_- P_{\nu_-} + D_+ Q_{\nu_+} + D_- Q_{\nu_-}, \\
\Phi_1
 & = &-\frac{2}{\mu^2} 
\{\partial_{\bar r}[(1-\bar r^2) \partial_{\bar r}\Phi_2]
+ 2\bar r \partial_{\bar r}\Phi_2 + \Phi_2\},
\end{eqnarray}
where indices $ \nu_\pm $ are determined by
%
\begin{equation}
\nu_\pm (\nu_\pm+1) = \lambda_\pm = \mu^2+1 \pm \sqrt{3\mu^2+1}.
\end{equation}
The first boundary condition is expanded around $ \bar r=1 $ as
%
\begin{eqnarray}
\Phi_1 + 2\Phi_2
 & \sim &2 \mu^{-2} (D_+ + D_-) (1-\bar r)^{-1}
   \nonumber \\
 && \quad - \mu^{-2} [(\lambda_+-1) D_+ + (\lambda_--1) D_-]
   \nonumber \\
 && \quad + \mathcal O(1-\bar r).
\end{eqnarray}
This means $ D_+ = D_- = 0 $, since $\lambda_+\ne\lambda_-$ for
$\mu^2\ne -1/3$.
Next we expand it around $ \bar r=-1 $, then
%
\begin{eqnarray}
\Phi_1 + 2\Phi_2
 & \sim &4 \pi^{-1} \mu^{-2} (C_+ \sin\nu_+\pi + C_- \sin\nu_-\pi)
 (1+\bar r)^{-1}
   \nonumber \\
 && \quad + 2 \pi^{-1} \mu^{-2}
     [(\lambda_+-1) C_+ \sin\nu_+\pi + (\lambda_--1) C_- \sin\nu_-\pi]
   \nonumber \\
 && \quad + \mathcal O(1+\bar r).
\end{eqnarray}
This implies that non-trivial solutions can exist if $ \nu_+ \in\mathbb Z$
or $ \nu_- \in \mathbb Z $.
With these choices of the parameters, we can confirm that the second
boundary condition in (\ref{bc_rbar2}) is satisfied.
In general we can choose non-negative $ \nu_\pm $'s so that these
conditions are explicitly written down as
%
\begin{equation}
\nu_\pm
 = \frac{-1+\sqrt{4\mu^2+5 \pm 4\sqrt{3\mu^2+1}}}{2} = 0,1,2,\ldots.
\end{equation}
Therefore the KK mass spectrum for scalar perturbation is obtained.
The case of $\mu^2\ne -1/3$ can be excluded\cite{Sendouda}.

\subsubsection{Numerical solution of KK modes}
\label{kkmode_scalar}

Here we obtain the first few KK modes of scalar type perturbations by
numerically solving the perturbed Einstein equation
(\ref{eqn:master-eq-scalar2}) and the junction condition 
(\ref{bc_rbar}).
We rewrite the system of two second order differential equations
(\ref{eqn:master-eq-scalar2}) to a system of four first order
differential equations by defining $\partial_{\bar r} \Phi_1$ and
$\partial_{\bar r} \Phi_2$ as well as $\Phi_1$ and
$\Phi_2$ as dependent variables.
Here, we take $M=51$.
We solve the problem while changing $\alpha$ from 1 to 0, each time with
a slightly different value of $\alpha$.
For $\alpha=1$, the analytic solutions presented in the previous
subsection are used as trial solutions.

\begin{figure}[tbp]
\leavevmode
\begin{center}
\includegraphics[width=7cm]{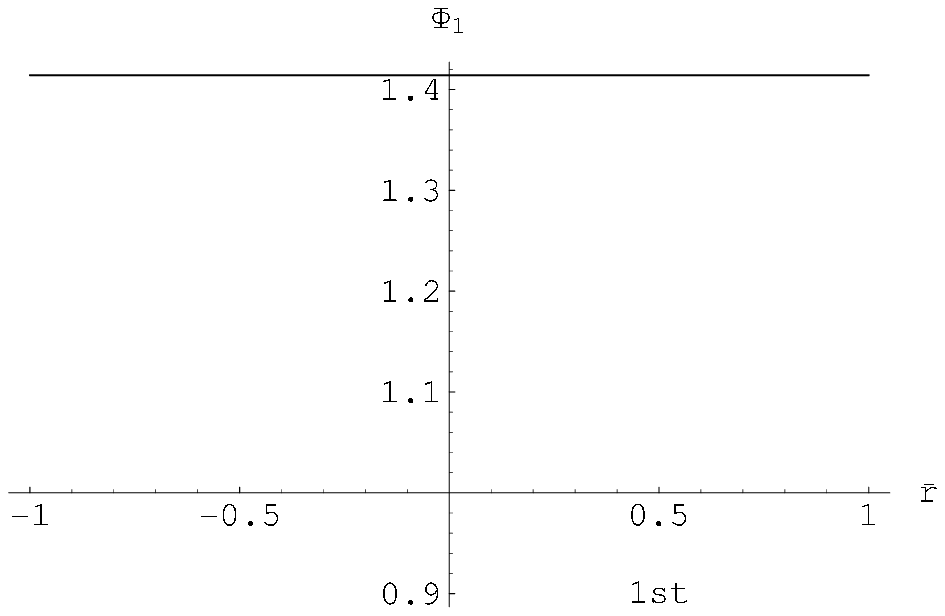}
\includegraphics[width=7cm]{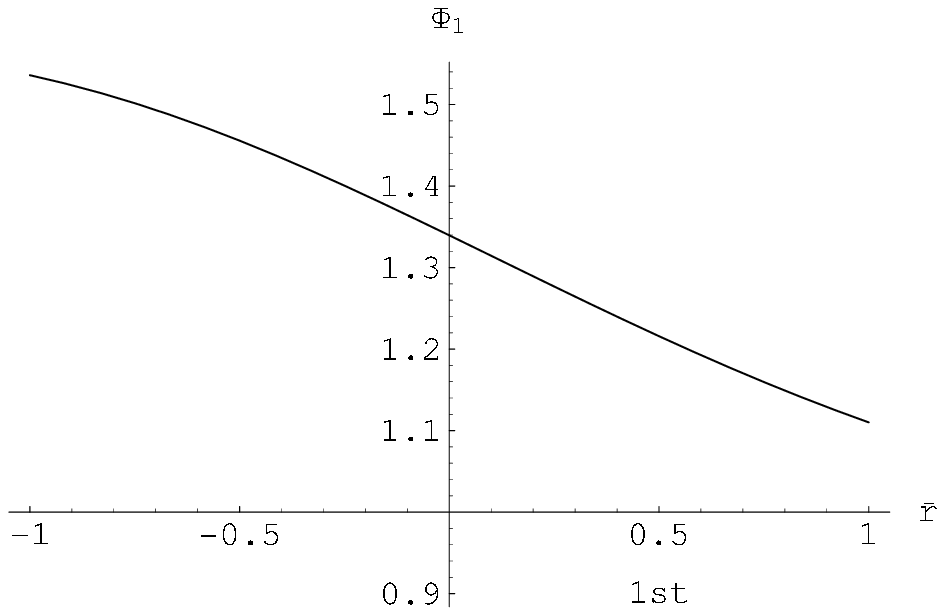}
\includegraphics[width=7cm]{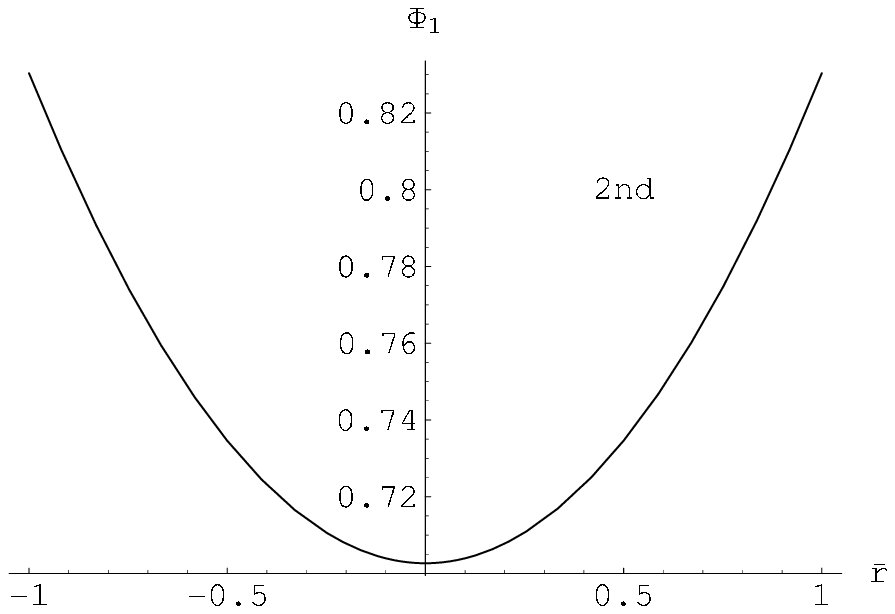}
\includegraphics[width=7cm]{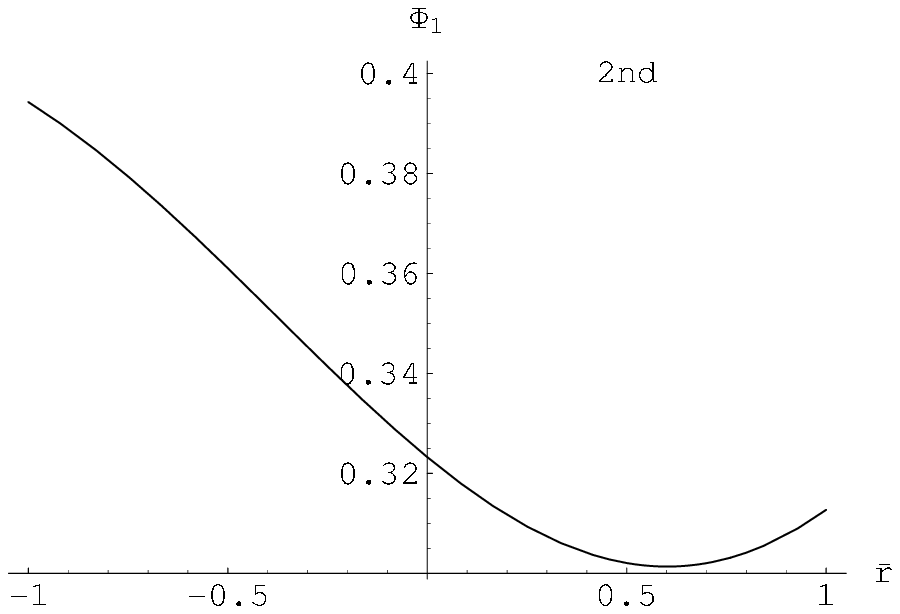}
\includegraphics[width=7cm]{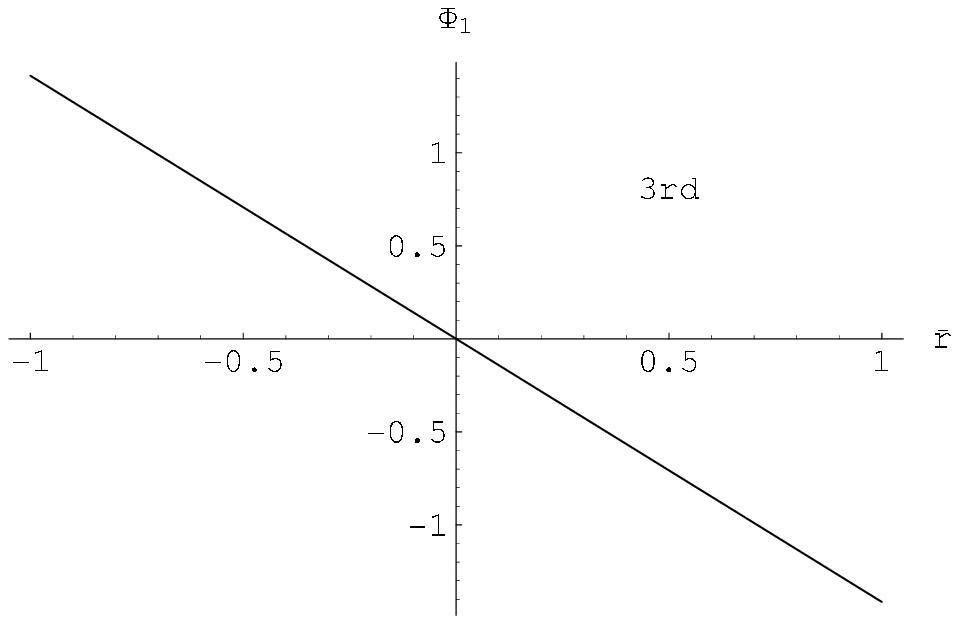}
\includegraphics[width=7cm]{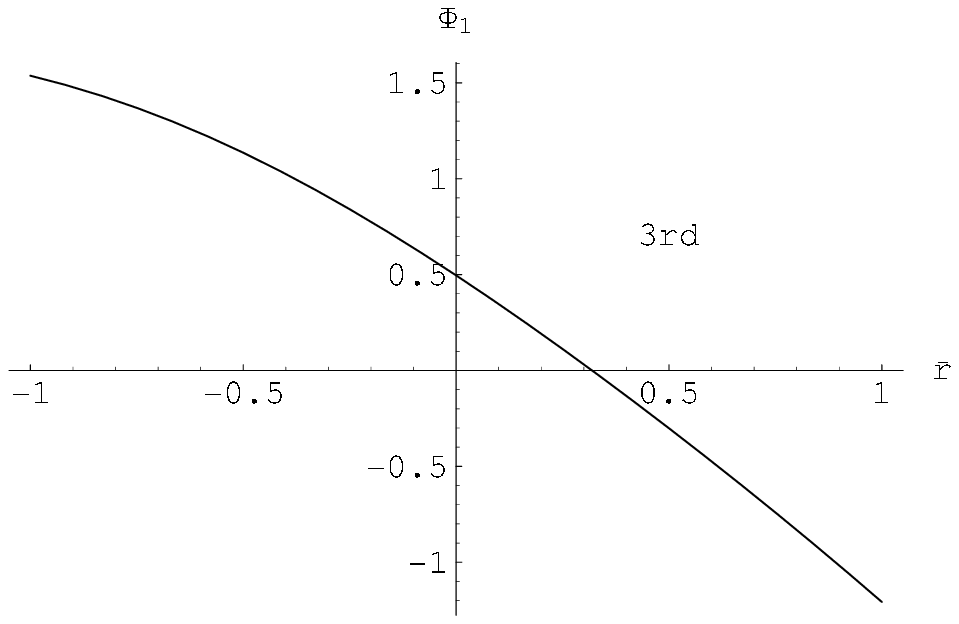}
\includegraphics[width=7cm]{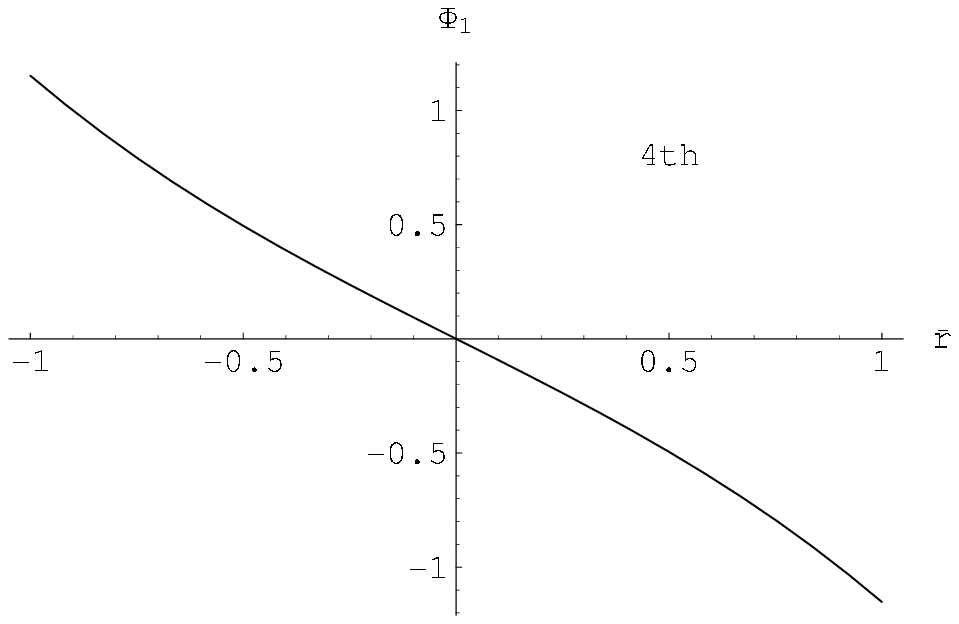}
\includegraphics[width=7cm]{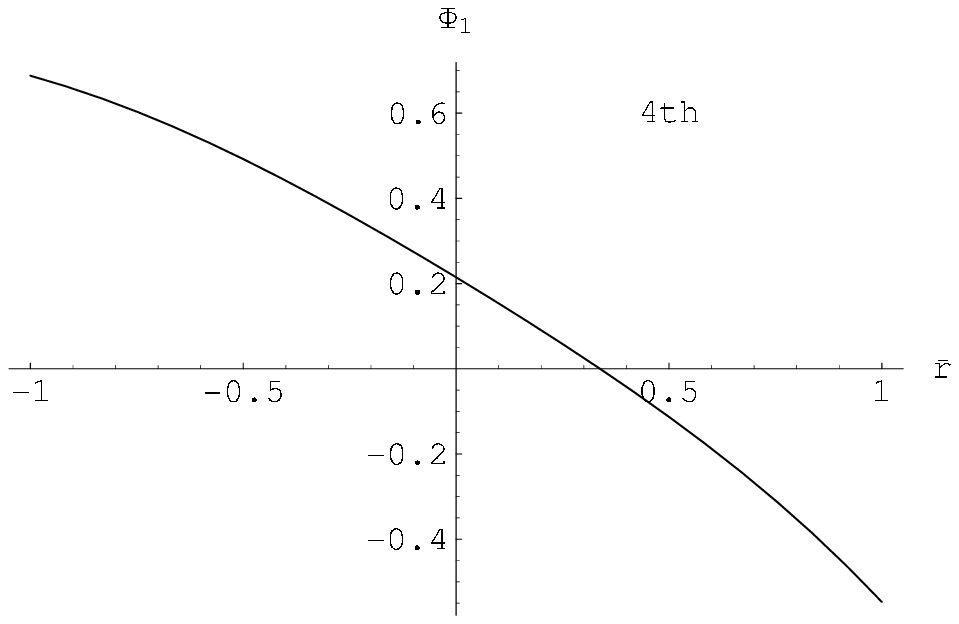}
\caption{The solution $\Phi_1$ of the first four KK modes for $\alpha=1$
 (left) and $\alpha=0.31$ (right).
The normalization is determined by using the generalized Klein-Gordon
 norm (see the text).
Number of points of the mesh is taken to be $51$.
}
\label{fig:scalar_1}
\end{center}
\end{figure}

\begin{figure}[tbp]
\leavevmode
\begin{center}
\includegraphics[width=7cm]{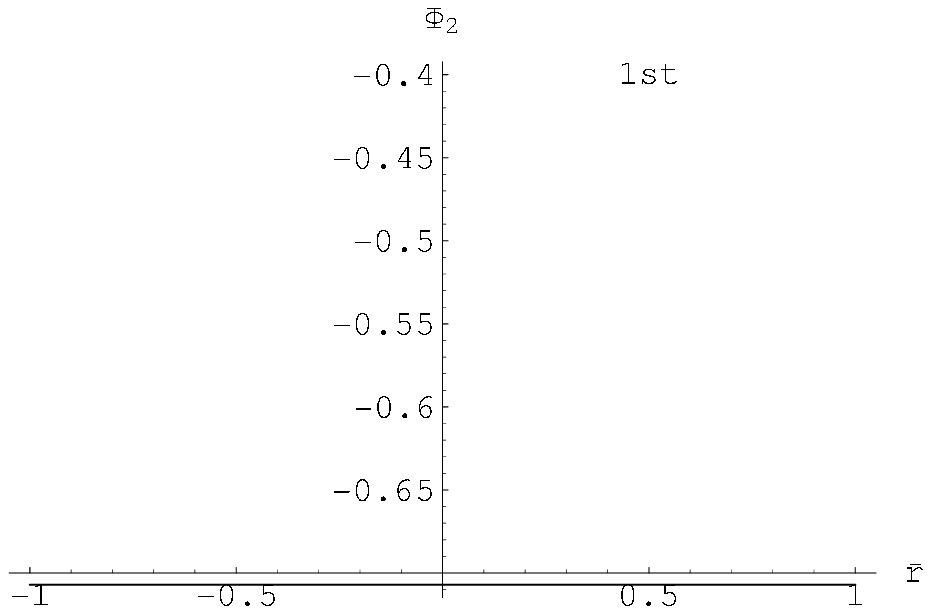}
\includegraphics[width=7cm]{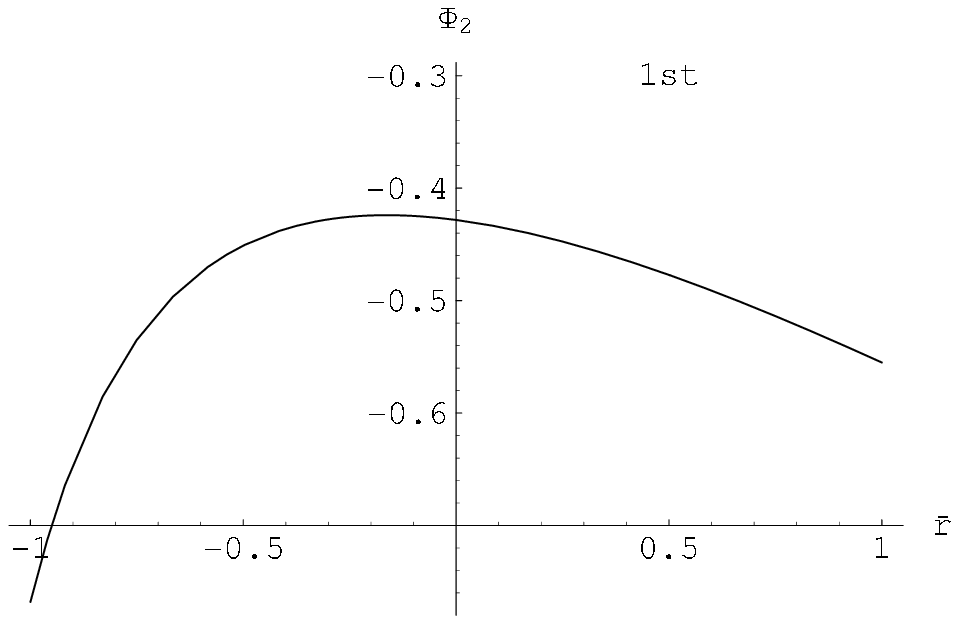}
\includegraphics[width=7cm]{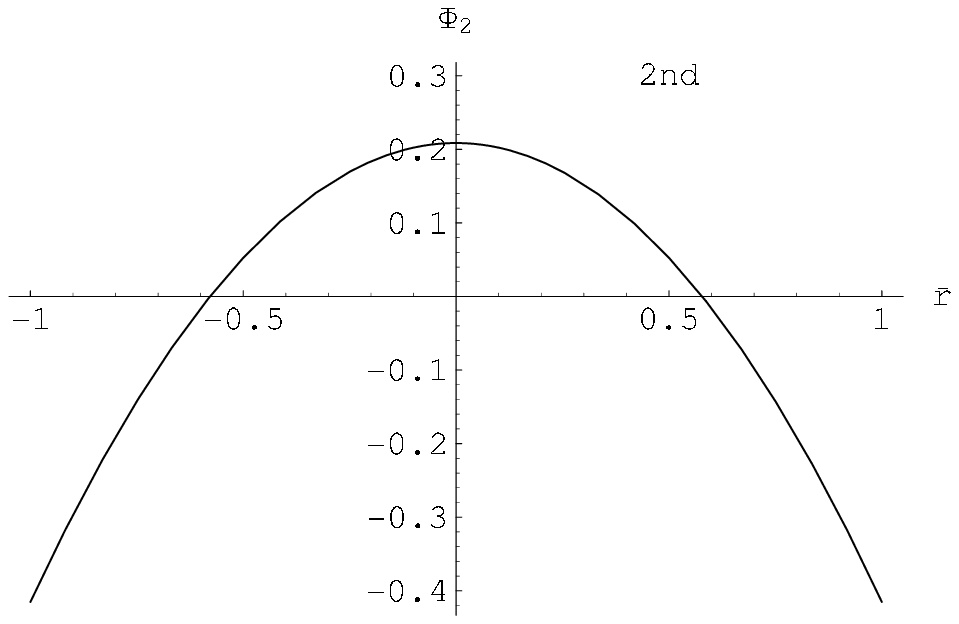}
\includegraphics[width=7cm]{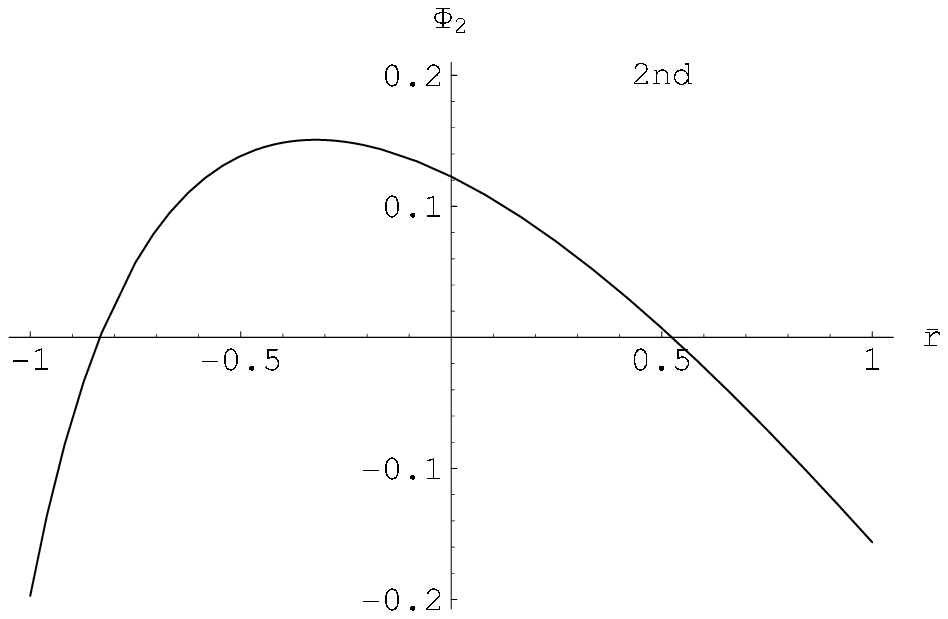}
\includegraphics[width=7cm]{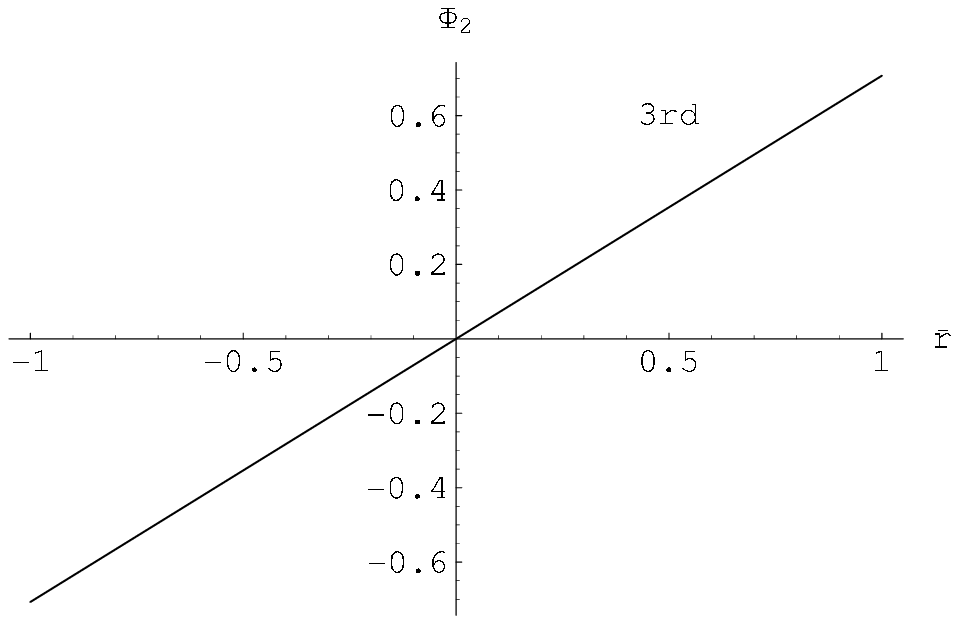}
\includegraphics[width=7cm]{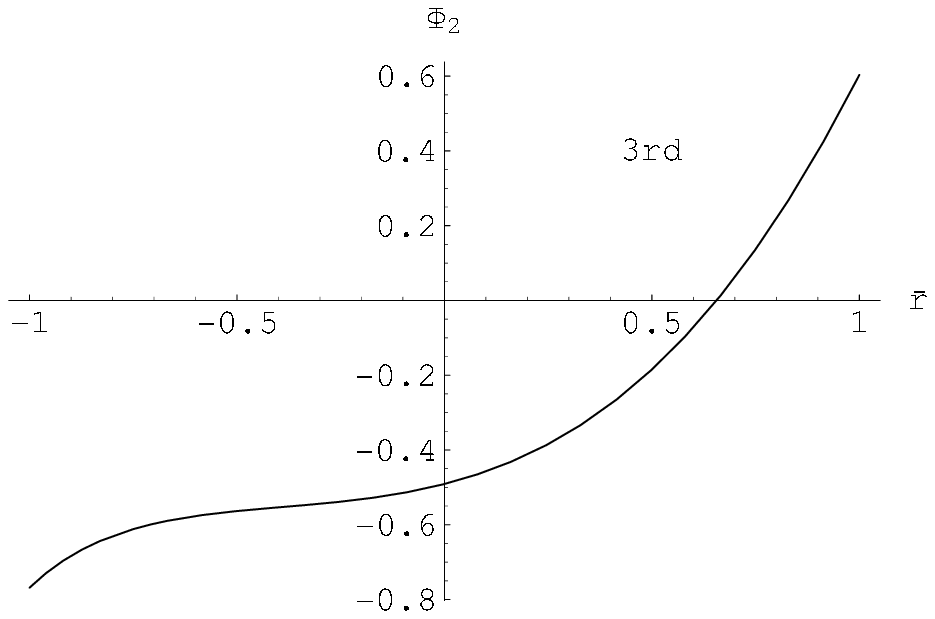}
\includegraphics[width=7cm]{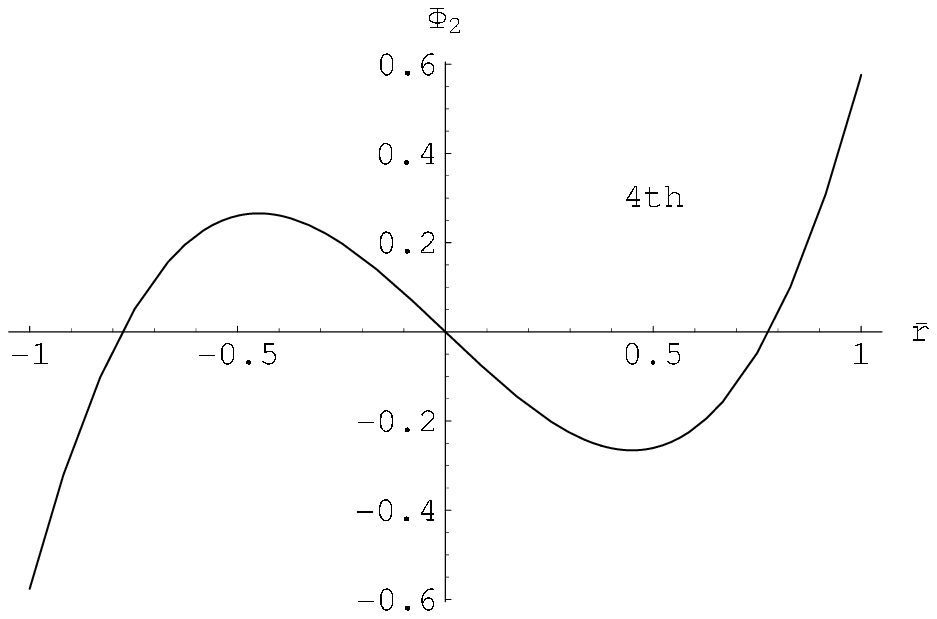}
\includegraphics[width=7cm]{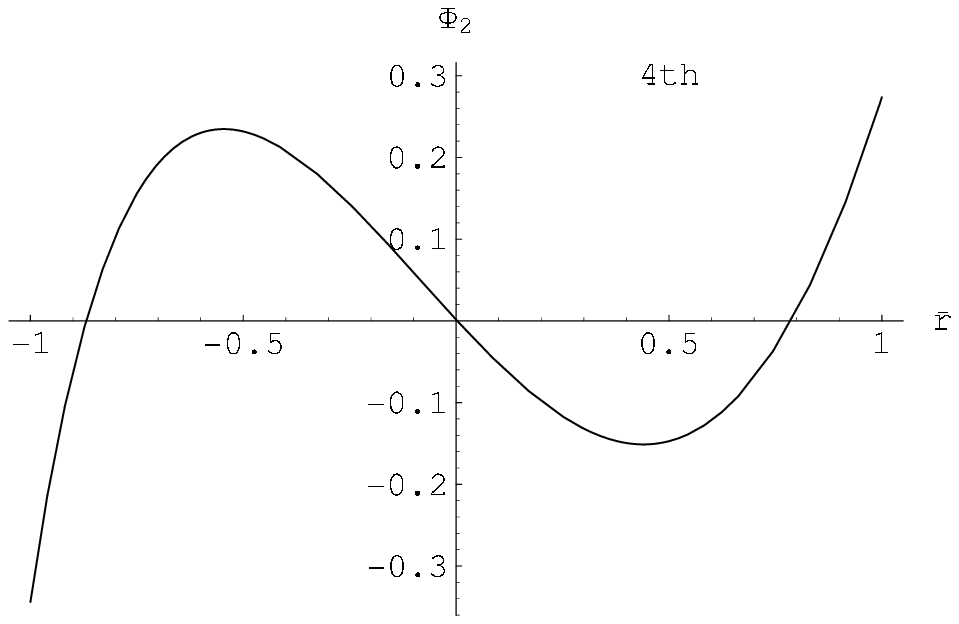}
\caption{The solution $\Phi_2$ of the first four KK modes for $\alpha=1$
 (left) and $\alpha=0.31$ (right).
The normalization is determined by using the generalized Klein-Gordon
 norm (see the text).
Number of points of the mesh is taken to be $51$.
}
\label{fig:scalar_2}
\end{center}
\end{figure}

Figure~\ref{fig:scalar_1} and ~\ref{fig:scalar_2} show the first four KK
mode solutions of $\Phi_1$ and $\Phi_2$ for $\alpha=1.0$ and $0.31$.
The normalization is determined by using the generalized Klein-Gordon
norm (\ref{KG_scalar}) given in the Appendix~\ref{KGnorm}.
In terms of the coordinate $(\bar r, \varphi)$, the equation
(\ref{KG_scalar}) is written as
%
\begin{eqnarray}
 (\tilde{\Phi},\tilde{\Psi})_{KG}
  & = & 
  M_6^4 (2\pi)^3(\omega_1+\omega_2)
  \delta^3({\bf k}_1-{\bf k}_2)e^{-i(\omega_1-\omega_2)t}
  \frac{\Delta\varphi r_+ r_-}{2\Lambda_6} \nonumber\\
 & & \times
  \left\{
  \int_{-1}^{1} 
  \frac{d\bar r}{2\left(8\beta_- (\beta_-\bar r +\beta_+)
  \partial_{\bar r}\bar f +24\beta_-^2\bar f
  +\Lambda_6 (\beta_-\bar r +\beta_+)^2\right)} \right. 
\nonumber\\ & &  \times 
  \left[
   \frac{1}{8}(\tilde m_1^2+\tilde m_2^2) (\beta_-\bar r +\beta_+)^2
   \Phi_1\Psi_1^*
   + \frac{(\beta_-\bar r +\beta_+)^4}{2}\partial_{\bar r}\bar f
   (\Psi_2^*\partial_{\bar r}\Phi_1+\Phi_2\partial_{\bar r}\Psi_1^*)
   \right.   \nonumber\\
 & & 
   + \frac{(\beta_-\bar r +\beta_+)^2}{4}\left(13\beta_-^2 \bar f
   +3\Lambda_6 \frac{(\beta_-\bar r +\beta_+)^2}{4}\right)\Phi_1\Psi_1^*
\nonumber\\ & &
   + (\beta_-\bar r +\beta_+)^2 \left(
   2\beta_- (\beta_-\bar r +\beta_+)\partial_{\bar r}\bar f
   +12\beta_-^2\bar f+ 3\Lambda_6 \frac{(\beta_-\bar r +\beta_+)^2}{4}
   \right)\Phi_2\Psi_2^*
   \nonumber\\
 & & \biggl.
   + \frac{(\beta_-\bar r +\beta_+)^2}{4\bar f}\biggl(
    (\beta_-\bar r +\beta_+)^2{\partial_{\bar r}\bar f}^2
    +10\beta_- (\beta_-\bar r +\beta_+) \bar f \partial_{\bar r}\bar f
    +24\beta_-^2 \bar f^2
\nonumber\\
 & &~~~
    +3\Lambda_6 \bar f \frac{(\beta_-\bar r +\beta_+)^2}{4}\biggr)
   (\Phi_1+2\Phi_2)(\Psi_1^*+2\Psi_2^*) \biggr]   \nonumber\\
 & & 
  \left.
  + \left[\frac{3(\beta_-\bar r + \beta_+)^3 \partial_{\bar r}\bar f
    \Phi_1\Psi_1^*}
   {16(8\beta_-\partial_{\bar r}\bar f +\Lambda_6 (\beta_-\bar r + \beta_+))}
  \right]_{\bar r=-1}^{\bar r=1}\right\}
\nonumber\\
  & \equiv &   M_6^4 (2\pi)^3(\omega_1+\omega_2)
  \delta^3({\bf k}_1-{\bf k}_2)e^{-i(\omega_1-\omega_2)t}
  \frac{\Delta\varphi r_+ r_-}{2\Lambda_6}
  \left(\Phi_1,\Phi_2\Big|\Psi_1,\Phi_2\right). 
\end{eqnarray}
We normalize the solution by
%
\begin{eqnarray}
\left(\Phi^{(n_1)}_1,\Phi^{(n_1)}_2\Big|
 \Phi^{(n_2)}_1,\Phi^{(n_2)}_2\right)
=\delta_{n_1 n_2},
\end{eqnarray}
where superscript $n_k$ means that $\Phi_1^{(n_k)}$ and 
$\Phi_2^{(n_k)}$ are the solutions of scalar perturbations
with eigenvalue $m_{n_k}^2$.

\begin{figure}[t]
\centerline{\includegraphics[width=12cm]{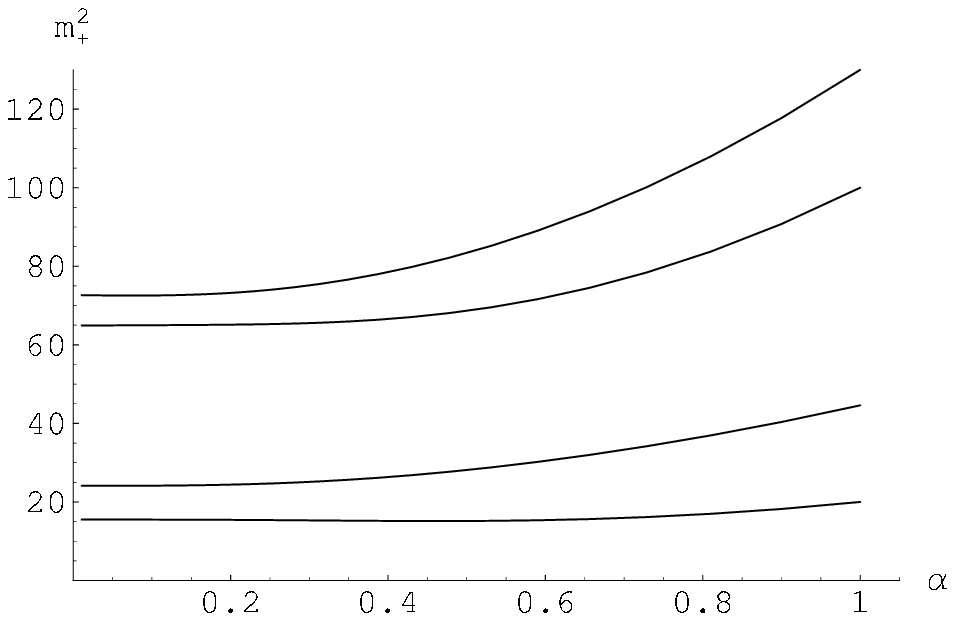}}
\caption{The spectrum of $m_+^2$ for scalar perturbations as a function
 of $\alpha$.}
\label{fig:sp_scalar}
\end{figure}

Finally, we show the spectrum of $m_+^2$ of the first four KK modes
as a function of $\alpha$ in figure~\ref{fig:sp_scalar}.
We find that $m_+^2$ is non-negative for the entire range of
$\alpha$.
Therefore, the background spacetime is dynamically
stable in the scalar-type sector.
The behavior of $m_+^2$ in $\alpha\to 0$ limit is also similar to the
cases of vector and tensor perturbations.
It remains finite in this limit, which is consistent with our previous
work~\cite{paper1}, where we obtained the energy scale at which the
correction to the effective Friedmann equation on the brane appears.

\section{Summary and Discussion}
\label{summary}

In this paper, we have considered the dynamical stability of the
six-dimensional brane world model with warped flux compactification
recently found by the authors.
This solution captures essential features of the warped flux
compactification, including warped geometry, compactification, a
magnetic flux, and one or two 3-brane(s).
For simplicity we have set the four-dimensional
cosmological constant to zero, and restricted to linear perturbations with
the axisymmetry corresponding to the rotation in the two-dimensional
bulk. We have expanded perturbations by scalar-, vector- and tensor-type
harmonics of the four-dimensional Minkowski spacetime and analyzed each
type separately.

The perturbations were labeled by its type and values of mass squared
$m^2 = -\eta^{\mu\nu}k_\mu k_\nu$.
To study the stability,
we have utilized the fact that for any perturbation type, the
perturbation equations are reduced to an eigenvalue problem with
eigenvalue $m^2$.
We regarded the background spacetime to be dynamically stable if
the spectrum of $m^2$ is non-negative.
This question was treated in each sector through the following steps.
\begin{itemize}
 \item [(i)] We have shown the reality of $m^2$. 
 \item [(ii)] We have rewritten the systems of the perturbation
       equations into a form which depends on background parameters 
       through just one parameter $\alpha$. ($0\leq\alpha\leq 1$)
 \item [(iii)] In the $\alpha\to 1$ limit we have analytically solved
       the perturbation equations and have shown that the spectrum of
       $m^2$ is non-negative. 
 \item [(iv)] We have numerically evaluated how each eigenvalue $m^2$
       changes as $\alpha$ moves from $1$ to $0$, and have shown that
       the spectrum remains non-negative throughout. 
\end{itemize}

For $\alpha<1$, we had to numerically solve the systems of the
perturbation equations many times, each time with a slightly different
values of $\alpha$.
In such a situation, relaxation method is useful, and thus we have
employed this method.
When $\alpha$ is slightly changed, the previous solution will be a good
initial guess, and relaxation works well.
Since the perturbations can be analytically solved in all the sectors
for $\alpha=1$, we have started the sequence of numerical computations
from $\alpha=1-\Delta\alpha$ by using the analytic solutions as the
initial guess, where $\Delta\alpha$ is a small positive number.

The mass squared that has physical meaning is
$m_\pm^2\equiv -r_\pm^{-2}\eta^{\mu\nu}k_\mu k_\nu$,
which is the one observed on the brane at $r=r_\pm$.
We have shown the spectra of $m_+^2$ as a function of $\alpha$ in each
sector, and found that $m_+^2$ is non-negative for the entire range of
$\alpha$, $\left[0,1\right]$.
Therefore, the background spacetime we consider is dynamical stable in
each sector.
We have found that there are zero modes only in the tensor sector,
corresponding to the four-dimensional gravitons.

Another remarkable feature of the spectra of $m_+^2$ is that they remain
finite in $\alpha\to 0$ limit, that is, $m_+^2\propto\alpha^0$ and
$m_-^2\propto\alpha^{-2}$ for $\alpha\to 0$ in all the sectors.
This result is consistent with our previous study~\cite{paper1}, where
we analyzed how the Hubble expansion rate $H_{\pm}$ on each brane
changes when the brane tension changes.
We also considered higher-order corrections of the effective Friedmann
equation with respect to $H_{\pm}$.
The result is that higher-order corrections appears when $H_{\pm}$ is
larger than a critical value $H_{*\pm}$.
For $\alpha\to 0$, we found that $H_{*\pm}$ behave as the equation
(\ref{Hbehave}).
Since the higher-order corrections are caused by the KK modes,
this energy scale corresponds to their mass squared.
Thus, the behavior of $m_\pm^2$ we obtained is consistent with our
previous result.

Having established the stability of the exact braneworld solution, there
are many subjects for future research, including  the recovery of 
$4$-dimensional linearized Einstein gravity and corrections to it, the
recovery of the $4$-dimensional Friedmann equation, properties of black
hole geometries~\cite{Kinoshita}, and so on.

\section*{Note added}

After this paper was submitted for publication, we were notified that
the exact solution considered by the authors had been already found in
\cite{GW2,LW} in advance of our previous paper~\cite{paper1}.

\section*{Acknowledgements}

We would like to thank David L. Wiltshire for kindly notifying us of
his previous study.
We are grateful to Hideaki Kudoh, Tetsuya Shiromizu and Takahiro
Tanaka for useful discussions and/or comments.
We also thank Katsuhiko Sato for continuing encouragement. 
HY and YS are supported in part by JSPS.
SM's work was partially supported by the JSPS Grant-in-Aid on Scientific
Research No.17740134.
SK is supported in part by a Grant-in-Aid for the 21st Century COE
Program ``Quantum Extreme Systems and Their Symmetries'' from the
Ministry of Education, Culture, Sports, Science, and Technology of
Japan. 

\appendix{Appendices}

\subsection{Harmonics in Minkowski spacetime}
\label{harmonics}

In this appendix we give definitions of scalar, vector and tensor
harmonics in an $n$-dimensional Minkowski spacetime. Throughout this 
appendix, $n$-dimensional coordinates are $x^{\mu}$
($\mu=0,1,\cdots,n-1$), $\eta_{\mu\nu}$ is the Minkowski metric, and all
indices are raised and lowered by the Minkowski metric and its inverse
$\eta^{\mu\nu}$. 

\subsubsection{Scalar harmonics}

The scalar harmonics are given by 
%
\begin{equation}
 Y = \exp(ik_{\rho}x^{\rho}),
\end{equation}
by which any function $f$ can be expanded as 
%
\begin{equation}
 f = \int dk\ c Y,
\end{equation}
where $c$ is a constant depending on $k$. Hereafter, $k$ and $dk$ are
abbreviations of $\{k^{\mu}\}$ ($\mu=0,1,\cdots,n-1$) and 
$\prod_{\mu=0}^{n-1}dk^{\mu}$, respectively. We omit $k$ in most cases. 

\subsubsection{Vector harmonics}

In general, any vector field $v_{\mu}$ can be decomposed as
%
\begin{equation}
 v_{\mu}=v_{(T)\mu}+\partial_{\mu} f ,
\end{equation}
where $f$ is a function and $v_{(T)\mu}$ is a transverse vector field:
%
\begin{equation}
 \partial^{\mu}v_{(T)\mu}=0 .
\end{equation}

Thus, the vector field $v_{\mu}$ can be expanded by using the scalar 
harmonics $Y$ and transverse vector harmonics $V_{(T)\mu}$ as 
%
\begin{equation}
 v_{\mu} = \int dk
  \left[c_{(T)}V_{(T)\mu}+c_{(L)}\partial_{\mu} Y\right].
	\label{eqn:dY+V}
\end{equation}
Here, $c_{(T)}$ and $c_{(L)}$ are constants depending on $k$, and the
transverse vector harmonics $V_{(T)\mu}$ are given by 
%
\begin{equation}
 V_{(T)\mu} = u_{\mu}\exp(ik_{\rho}x^{\rho}),
\end{equation}
where the constant vector $u_{\mu}$ satisfies the following condition. 
%
\begin{equation}
 k^{\mu}u_{\mu}=0
\end{equation}
for $k^{\mu}k_{\mu}\ne 0$, and 
%
\begin{eqnarray}
 k^{\mu}u_{\mu} & = & 0, \nonumber\\
 \tau^{\mu}u_{\mu} & = & 0
  \label{eqn:u-for-k^2=0}
\end{eqnarray}
for non-vanishing $k_{\mu}$ satisfying $k^{\mu}k_{\mu}=0$, where
$\tau^{\mu}$ is an arbitrary constant timelike vector. 
For $k_{\mu}=0$,
the constant vector $u^{\mu}$ does not need to satisfy any of the above
conditions. For the special case $k^{\mu}k_{\mu}=0$, the second
condition in (\ref{eqn:u-for-k^2=0}) can be imposed by redefinition of
$c_{(L)}$. Actually this condition is necessary to eliminate
redundancy. Note that the number of independent vectors satisfying the
above condition is $n-1$ for $k^{\mu}k_{\mu}\ne 0$ and $n-2$ for
$k^{\mu}k_{\mu}=0$ and that these numbers are equal to the numbers of
physical degrees of freedom for massive and massless spin-$1$ fields in
$n$-dimensions, respectively.

Because of the expansion (\ref{eqn:dY+V}), it is convenient to define
longitudinal vector harmonics $V_{(L)\mu}$ by
%
\begin{equation}
 V_{(L)\mu} \equiv \partial_{\mu} Y = ik_{\mu}Y. 
\end{equation}

\subsubsection{Tensor harmonics}

In general, a symmetric second-rank tensor field $t_{\mu\nu}$ can be
decomposed as
%
\begin{equation}
 t_{\mu\nu}=t_{(T)\mu\nu} + \partial_{\mu}v_{\nu}+\partial_{\nu}v_{\mu} 
  + f\eta_{\mu\nu},
\end{equation}
where $f$ is a function, $v_{\mu}$ is a vector field and $t_{(T)\mu\nu}$
is a transverse traceless symmetric tensor field:
%
\begin{eqnarray}
 t_{(T)\mu}^\mu & = & 0,\nonumber\\
 \partial^{\mu} t_{(T)\mu\nu} & = &0.
	\label{eqn:trasverse-traceless}
\end{eqnarray}

Thus, the tensor field $t_{\mu\nu}$ can be expanded by using the scalar
harmonics $Y$, the vector harmonics $V_{(T)}$ and $V_{(L)}$, and
transverse traceless tensor harmonics $T_{(T)}$ as 
%
\begin{eqnarray}
 t_{\mu\nu} & = & \int dk\left[
	c_{(T)}T_{(T)\mu\nu}+c_{(LT)}
	(\partial_{\mu}V_{(T)\nu}+\partial_{\nu}V_{(T)\mu})\right.
	\nonumber\\
 & &	\left.
	 + c_{(LL)}(\partial_{\mu}V_{(L)\nu}+\partial_{\nu}V_{(L)\mu})
	+ \tilde{c}_{(Y)}Y\eta_{\mu\nu}\right].
	\label{eqn:dV+T}
\end{eqnarray}
Here, $c_{(T)}$, $c_{(LT)}$, $c_{(LL)}$, and $\tilde{c}_{(Y)}$ are
constants depending on $k$, and the transverse traceless tensor
harmonics $T_{(T)}$ are given by 
%
\begin{equation}
 T_{(T)\mu\nu} = s_{\mu\nu}\exp(ik_{\rho}x^{\rho}),
\end{equation}
where the constant symmetric second-rank tensor $s_{\mu\nu}$ satisfies
the following condition. 
%
\begin{eqnarray}
 k^{\mu}s_{\mu\nu} & = & 0, \nonumber\\
 s^{\mu}_{\mu} & = & 0
\end{eqnarray}
for $k^{\mu}k_{\mu}\ne 0$, and 
%
\begin{eqnarray}
 k^{\mu}s_{\mu\nu} & = & 0, \nonumber\\
 s^{\mu}_{\mu} & = & 0, \nonumber\\
 \tau^{\mu}s_{\mu\nu} & = & 0
  \label{eqn:s-for-k^2=0}
\end{eqnarray}
for non-vanishing $k_{\mu}$ satisfying $k^{\mu}k_{\mu}=0$, where
$\tau^{\mu}$ is an arbitrary constant timelike vector. 
For $k_{\mu}=0$, the constant tensor $s_{\mu\nu}$ does not need to
satisfy any of the above conditions. For the special case
$k^{\mu}k_{\mu}=0$, the last condition in (\ref{eqn:s-for-k^2=0}) can be
imposed by redefinition of $c_{(LT)}$, $c_{(LL)}$ and
$\tilde{c}_{(Y)}$. Actually this condition is 
necessary to eliminate redundancy. Note that the number of independent
symmetric second-rank tensors satisfying the above conditions is
$(n+1)(n-2)/2$ for $k^{\mu}k_{\mu}\ne 0$ and $n(n-3)/2$ for 
$k^{\mu}k_{\mu}=0$ and that these numbers are equal to numbers of
physical degrees of freedom for massive and massless spin-$2$ fields in
$n$-dimensions, respectively.

Because of the expansion (\ref{eqn:dV+T}), it is convenient to define
tensor harmonics $T_{(LT)}$, $T_{(LL)}$, and $T_{(Y)}$ by 
%
\begin{eqnarray}
 T_{(LT)\mu\nu} & \equiv & \partial_{\mu}V_{(T)\nu}
  +\partial_{\nu}V_{(T)\mu}, \nonumber\\
 & = & i(u_{\mu}k_{\nu}+u_{\nu}k_{\mu})Y, \nonumber\\
 T_{(LL)\mu\nu} & \equiv & \partial_{\mu}V_{(L)\nu}
  +\partial_{\nu}V_{(L)\mu}
	-\frac{2}{n}\eta_{\mu\nu}\partial^{\rho}V_{(L)\rho} \nonumber\\
 & = & \left(-2k_{\mu}k_{\nu}+\frac{2}{n}k^{\rho}k_{\rho}
	\eta_{\mu\nu}\right)Y,  \nonumber\\ 
 T_{(Y)\mu\nu} & \equiv & \eta_{\mu\nu}Y. 
\end{eqnarray}

\subsection{Gauge transformation of the perturbations}
\label{gauge}

In this appendix we give gauge transformations of the perturbations of
the metric and the $U(1)$ gauge field.
The coordinate gauge transformation is of the form
%
\begin{equation}
x^M \to x^M + \bar\xi^M.
\end{equation}
There is another kind of gauge transformation.
The perturbations of the field strength $\delta F_{MN}$ are not changed
under gauge transformation of the gauge field,
%
\begin{equation}
\delta A_M \to \delta A_M + \partial_M \bar\zeta.
\end{equation}
The gauge parameters $\bar\xi^M$ and $\bar\zeta$ can be expanded by the
scalar and the vector harmonics as
%
\begin{eqnarray}
&&\bar\xi_M dx^M = \left( \xi_{(T)}V_{(T)\mu}
+ \xi_{(L)}V_{(L)\mu}\right)dx^\mu + \xi_r Y dr + \xi_{\phi}Yd\phi,
\\&&
\bar\zeta = \zeta Y,
\end{eqnarray}
where $\xi_{(T,L)}$, $\xi_r$, $\xi_{\phi}$ and $\zeta$ are supposed to
depend only on $r$.
Under the above gauge transformation, the perturbation variables
transform as
%
\begin{eqnarray}
&&\bar h_{(T)} = h_{(T)},
\\&&
\bar h_{(LT)} = h_{(LT)} - \xi_{(T)},
   \label{eqn:gauge-tr-hLT}
\\&&
\bar h_{(LL)} = h_{(LL)} - \xi_{(L)},
\\&&
\bar h_{(Y)} = h_{(Y)} -2rf\xi_r + \frac{1}{2}k^\mu k_\mu \xi_{(L)},
\\&&
\bar h_{(T)r} = h_{(T)r} + \frac{2}{r}\xi_{(T)} - \xi_{(T)}',
   \label{eqn:gauge-tr-hTr}
\\&&
\bar h_{(L)r} = h_{(L)r} + \frac{2}{r}\xi_{(L)} - \xi_{(L)}' -\xi_r,
\\&&
\bar h_{(L)\phi} = h_{(L)\phi} - \xi_{\phi},
\\&&
\bar h_{rr} = h_{rr} -2\xi_r' -\frac{f'}{f}\xi_r,
\\&&
\bar h_{r\phi} = h_{r\phi} -\xi_{\phi}' + \frac{f'}{f}\xi_{\phi},
\\&&
\bar h_{\phi\phi} = h_{\phi\phi} -f f' \xi_r,
\\&&
\bar a_{(T)} = a_{(T)},
\\&&
\bar a_{(L)} = a_{(L)} + \zeta - \frac{A}{f}\xi_{\phi},
\\&&
\bar a_r = a_r + \partial_r \zeta -
\frac{A}{f}\left(\xi_{\phi}'-\frac{f'}{f}\xi_{\phi}\right), 
\\&&
\bar a_\phi = a_\phi - fA'\xi_r.
\end{eqnarray}

Finally we summarize the gauge conditions employed in this paper.
Of course, tensor type perturbations are gauge invariant from the
beginning.
For vector perturbations, we set $\bar h_{(LT)}=0$ by taking
$\xi_{(T)}=h_{(LT)}$.
The master equations for scalar perturbations in the main part of 
section~\ref{scalar} are derived in the analog of the Longitudinal
gauge where $\bar h_{(LL)}=\bar h_{(L)r}=\bar{h}_{r\phi}=0$.
This can be obtained by taking a gauge
%
\begin{eqnarray}
&&\xi_{(L)}=h_{(LL)},
\\&&
\xi_r = h_{(L)r}+\frac{2}{r}h_{(LL)} - h_{(LL)}',
\\&&
\xi_{\phi} = \tilde{C}f(r) + f(r)\int^r dr'\frac{h_{r\phi}(r')}{f(r')}, 
\label{residual-gauge-freedom}
\end{eqnarray}
where $\tilde{C}$ is an arbitrary constant representing the residual
gauge freedom. We use this residual gauge freedom to set $C=0$ in
subsection~\ref{basic}.
In subsection~\ref{reality}, we show the reality of $m^2$ for scalar
perturbations by adopting another gauge.
This gauge corresponds to $\bar h_{(LL)}=0$, 
$3f\bar h_{(Y)}+r^2 \bar h_{\phi\phi}=0$ and $\bar{h}_{r\phi}=0$, which
can be set by 
%
\begin{eqnarray}
&&\xi_{(L)}=h_{(LL)},
\\&&
\xi_r = \frac{3fh_{(Y)}+r^2 h_{\phi\phi}+3k^\mu k_\mu h_{(LL)}}
{rf\left(6f+rf'\right)},
\\&&
\xi_{\phi} = \tilde{C}f(r) + f(r)\int^r dr'\frac{h_{r\phi}(r')}{f(r')}, 
\end{eqnarray}
where $\tilde{C}$ is again an arbitrary constant representing the
residual gauge freedom. 
For the U(1) gauge field, we set $\bar a_{(L)}=0$ by $\zeta=a_{(L)}$
both in the main part of section~\ref{scalar} and in
subsection~\ref{reality}.

\subsection{Relaxation method}
\label{relax}

In the following we explain relaxation method in detail~\cite{recipes}.
First of all, we rewrite a system of second-order differential equations
to a system of first-order differential equations of the form
%
\begin{equation}
\frac{dy_i \left(\bar r\right)}{d\bar r} = g_i\left(\bar
	    r,y_1,y_2,\cdots,y_N,\tilde m^2\right)~~~
\left(i=1,2,\cdots,N\right),
\label{four_ODE}
\end{equation}
where $y_i$ denotes one of $N$ dependent functions.
For example, $y_1=h$ and $y_2=\partial_{\bar r}h$ for tensor
perturbations.
When we numerically solve the equations, one of the dependent
functions has to be fixed at some $\bar r$ in an arbitrary manner.
(When we plot the solution, we use another normalization.)
Thus, these dependent functions have to satisfy $N+1$ boundary
conditions instead of just $N$.
The problem is overdetermined and in general there is no solution for
arbitrary values of $\tilde m^2$.
For certain special values of $\tilde m^2$, the Eq.(\ref{four_ODE}) does
have a solution.
Such $\tilde m^2$ is the eigenvalue.
It is convenient to reduce this problem to the standard case by
introducing a new dependent variable
%
\begin{equation}
y_{N+1} \equiv \tilde m^2
\end{equation}
and another differential equation
%
\begin{equation}
\frac{dy_{N+1}}{d\bar r}=0.
\end{equation}

Next, we replace the differential equations with finite-difference
equations.
We first define a mesh by a set of $k=1,2,\cdots,M$ points at which we
supply values for the independent variable $\bar r_k$.
In particular, $\bar r_1\left(=-1\right)$ is the initial boundary, and
$\bar r_M\left(=1\right)$ is the final boundary.
We use the notation ${\bf y}_k$ to refer to the entire set of dependent
variables $y_1,y_2,\cdots,y_{N+1}$ at point $\bar r_k$.
At an arbitrary point $k$ in the middle of the mesh, we approximate the
set of five differential equations by algebraic relations of the form
%
\begin{equation}
0={\bf E}_k \equiv {\bf y}_k - {\bf y}_{k-1} 
-\left({\bar r}_k - {\bar r}_{k-1}\right)
{\bf g}_k \left({\bar r}_k,{\bar r}_{k-1},{\bf y}_k,{\bf y}_{k-1}\right),
~~~k=2,3,\cdots,M.
\label{finite1}
\end{equation}
The finite-difference equations labeled by ${\bf E}_k$ provide a total
of $\left(N+1\right)\left( M-1 \right)$ equations for the
$\left(N+1\right)M$ unknowns.
The remaining $N+1$ equations come from the boundary conditions.
At the first boundary we have
%
\begin{equation}
0={\bf E}_1 \equiv {\bf B}\left({\bar r}_1,{\bf y}_1\right)
\label{finite2}
\end{equation}
while at the second boundary
%
\begin{equation}
0={\bf E}_{M+1} \equiv {\bf C}\left({\bar r}_M,{\bf y}_M\right).
\label{finite3}
\end{equation}
The vectors ${\bf B}$ and ${\bf C}$ have only $n_2$ and $n_1=N+1-n_2$
nonzero components respectively, corresponding to number of the boundary
conditions.

The solution of the Eq.(\ref{finite1}), (\ref{finite2}), and
(\ref{finite3}) consists of a set of variables ${\bf y}_k$ at the $M$
points $\bar r_k$.
The algorithm we now describe requires an initial guess for ${\bf y}_k$.
We then determine increments $\Delta {\bf y}_k$ such that
${\bf y}_k + \Delta {\bf y}_k$ is an improved approximation to the
solution.
Equations for the increments are developed by expanding the
Eq.(\ref{finite1}) in first-order Taylor series with respect to
$\Delta {\bf y}_k$.
At an interior point, $k=2,3,\cdots,M$ this gives
%
\begin{eqnarray}
&&0={\bf E}_k\left({\bf y}_k + \Delta {\bf y}_k,
{\bf y}_{k-1} + \Delta {\bf y}_{k-1}\right) \sim
{\bf E}_k\left({\bf y}_k, {\bf y}_{k-1}\right)
\nonumber\\&&~~~~
+\sum_{n=1}^{N+1} \frac{\partial {\bf E}_k}{\partial y_{n,k-1}}
\Delta y_{n,k-1}
+\sum_{n=1}^{N+1} \frac{\partial {\bf E}_k}{\partial y_{n,k}}
\Delta y_{n,k},
\label{eq_delta1}
\end{eqnarray}
where $y_{n,k}$ is the value of $y_n$ at the point $\bar r_k$.
This provides $\left(N+1\right)\left(M-1\right)$ equations for
$\Delta {\bf y}_k$.
Similarly, the algebraic relations at the boundaries can be expanded in
a first-order Taylor series for increments that improve the solution.
Since ${\bf E}_1$ depends only on ${\bf y}_1$, we find at the first
boundary:
%
\begin{eqnarray}
0={\bf E}_1 + \sum_{n=1}^{N+1}
\frac{\partial {\bf E}_1}{\partial y_{n,1}} \Delta y_{n,1}.
\label{eq_delta2}
\end{eqnarray}
At the second boundary,
%
\begin{eqnarray}
0={\bf E}_{M+1} + \sum_{n=1}^{N+1}
\frac{\partial {\bf E}_{M+1}}{\partial y_{n,M}} \Delta y_{n,M}.
\label{eq_delta3}
\end{eqnarray}
We again note that the Eq.(\ref{eq_delta2}) and (\ref{eq_delta3}) have
only $n_2$ and $n_1=N+1-n_2$ nonzero components.
We thus have a set of $\left(N+1\right)M$ linear equations to be solved
for the corrections $\Delta {\bf y}_k$, iterating until the corrections
are sufficiently small.

\subsection{The generalized Klein-Gordon norm}
\label{KGnorm}

To normalize mode functions we use the generalized Klein-Gordon norm,
whose definition can be easily read off from the kinetic term in the
effective action for the corresponding physical degree of freedom. See,
for example, Appendix A of ref.~\cite{Mukohyama1999} for the definition
and the motivation of the generalized Klein-Gordon norm. 

\subsubsection{Tensor perturbation}

For tensor perturbation, 
%
\begin{eqnarray}
 ds_6^2 & = & r^2 
  \left[\eta_{\mu\nu} + h_{\mu\nu}\right]
  dx^{\mu}dx^{\nu} + \frac{dr^2}{f}
  + fd\phi^2,\nonumber\\ 
 A_Mdx^M & = & Ad\phi,
\end{eqnarray}
where $h_{\mu\nu}$ is a symmetric, transverse and traceless
four-dimensional tensor depending on ($x^{\mu}$, $r$), the bulk action is
expanded up to the second order in perturbation as 
%
\begin{eqnarray}
 I_6 & = &
  \frac{M_6^4}{2}\int d^6x\sqrt{-g}
  \left(R-2\Lambda_6-\frac{1}{2}F^{MN}F_{MN}\right)\nonumber\\
  & = & 
  -\frac{M_6^4\Delta\phi}{16} 
   \int d^4 x\int dr r^2\eta^{\rho\rho'}\eta^{\sigma\sigma'}
   \left[\eta^{\mu\nu}\partial_{\mu}h_{\rho\sigma}
       \partial_{\nu}h_{\rho'\sigma'}
       +fr^2\partial_rh_{\rho\sigma}\partial_rh_{\rho'\sigma'}\right]. 
\end{eqnarray}
Here, we have not written down the boundary term since it does not
change the definition of the generalized Klein-Gordon norm. It is easy
to check that the correct equation of motion is derived from this
action.

From this form of the action we can read off the generalized
Klein-Gordon norm as
%
\begin{equation}
 (\Phi,\Psi)_{KG} \equiv -i \frac{M_6^4\Delta\phi}{8}
   \int d^3{\bf x}\int dr r^2  \eta^{\mu\mu'}\eta^{\nu\nu'}
  \left(\Phi_{\mu\nu}\partial_t\Psi^*_{\mu'\nu'}
   -\Psi^*_{\mu\nu}\partial_t\Phi_{\mu'\nu'}\right).
\end{equation}

\subsubsection{Vector perturbation}

For vector perturbation, after fixing the gauge freedom ($h_{(LT)}=0$)
and using the corresponding constraint equation (the $(LT)$ component
of the Einstein equation), the metric and the $U(1)$ field in the
linearized level are written as
%
\begin{eqnarray}
 ds_6^2 & = & r^2 \eta_{\mu\nu}dx^{\mu}dx^{\nu}
  + 2  \left[\frac{h_{r\mu}}{r^2f}dr + h_{\phi\mu}d\phi\right]dx^{\mu}
   + \frac{dr^2}{f} + fd\phi^2,\nonumber\\
 A_Mdx^M & = & 
  a_{\mu}dx^{\mu} +  Ad\phi,
\end{eqnarray}
where $a_{\mu}$ and $h_{\phi\mu}$ are transverse four-dimensional vectors
depending on ($x^{\mu}$, $r$) and $h_{r\mu}$ is a transverse
four-dimensional vector depending only on $x^{\mu}$. The bulk action is 
expanded up to the second order in perturbation as 
%
\begin{eqnarray}
 I_6 & = &
  \frac{M_6^4}{2}\int d^6x\sqrt{-g}
  \left(R-2\Lambda_6-\frac{1}{2}F^{MN}F_{MN}\right)\nonumber\\
  & = & 
   \frac{M_6^4\Delta\phi}{4}\int d^4 x
  \int dr L,
\end{eqnarray}
where
%
\begin{equation}
 L = -\eta^{\rho\sigma}\left[
  2\eta^{\mu\nu}\partial_{\mu}a_{\rho}\partial_{\nu}a_{\sigma}
  + \frac{1}{f}\eta^{\mu\nu}\partial_{\mu}h_{\rho}\partial_{\nu}h_{\sigma} 
  + 2fr^2\partial_ra_{\rho}\partial_ra_{\sigma}
  - 4 r^2A'h_{\rho}\partial_ra_{\sigma}
  + r^2\partial_rh_{\rho}\partial_rh_{\sigma}
  + 6h_{\rho}h_{\sigma} \right].
\end{equation}
We have not written down the boundary term since it does not change the
definition of the generalized Klein-Gordon norm.  It is easy to check
that the correct equations of motion are derived from this action.

From this form of the action we can read off the generalized
Klein-Gordon norm as
%
\begin{equation}
 (\tilde{\Phi},\tilde{\Psi})_{KG} \equiv -i \frac{M_6^4\Delta\phi}{2}
  \int d^3{\bf x}\int dr \eta^{\mu\nu}
  \left[\left(\tilde{\Phi}_{1\mu}\partial_t\tilde{\Psi}_{1\nu}^*
	 -\tilde{\Psi}^*_{1\mu}\partial_t\tilde{\Phi}_{1\nu}\right)
   +\frac{r^4}{f}\left(\tilde{\Phi}_{2\mu}\partial_t\tilde{\Psi}_{2\nu}^* 
   -\tilde{\Psi}^*_{2\mu}\partial_t\tilde{\Phi}_{2\nu}\right)\right],
\end{equation}
where the solutions $\tilde{\Phi}$ and $\tilde{\Psi}$ are specified by
($\tilde{\Phi}_{1\mu}$, $\tilde{\Phi}_{2\mu}$) and ($\tilde{\Psi}_{1\mu}$,
$\tilde{\Psi}_{2\mu}$), respectively, as 
%
\begin{eqnarray}
 \tilde{\Phi} : &  & a_{\mu} = \frac{\tilde{\Phi}_{1\mu}}{\sqrt{2}}, \quad
  h_{\mu} = r^2\tilde{\Phi}_{2\mu}, \nonumber\\
 \tilde{\Psi} : &  & a_{\mu} = \frac{\tilde{\Psi}_{1\mu}}{\sqrt{2}}, \quad 
  h_{\mu} = r^2\tilde{\Psi}_{2\mu}.
\end{eqnarray}

\subsubsection{Scalar perturbation}

For scalar perturbation, after fixing the gauge freedom
($h_{(LL)}=h_{(L)r}=h_{(L)\phi}=h_{r\phi}=a_{(L)}=0$) and using the
corresponding constraint equations (the $(LL)$, $(L)r$, $(L)\phi$ and
$r\phi$ components of the Einstein equation and the $(L)$ component of
the Maxwell equation), the metric and the $U(1)$ field in the linearized
level are written as 
%
\begin{eqnarray}
 ds_6^2 & = & r^2(1+\tilde{\Phi}_2)
  \eta_{\mu\nu}dx^{\mu}dx^{\nu}
  + \left[1+(\tilde{\Phi}_1+\tilde{\Phi}_2)\right]\frac{dr^2}{f}
  + \left[1-(\tilde{\Phi}_1+3\tilde{\Phi}_2)\right]fd\phi^2,\nonumber\\
 A_Mdx^M & = & 
  \left\{A+ \frac{1}{A'}
   \left[\frac{1}{2r^2}(fr^2\tilde{\Phi}_1)'+ f'\tilde{\Phi}_2\right]
			 \right\}d\phi,
\end{eqnarray}
where $\tilde{\Phi}_1$ and $\tilde{\Phi}_2$ are functions of ($x^{\mu}$, $r$).
The bulk action is expanded up to the second order in perturbation as
%
\begin{eqnarray}
 I_6 & = &
  \frac{M_6^4}{2}\int d^6x\sqrt{-g}
  \left(R-2\Lambda_6-\frac{1}{2}F^{MN}F_{MN}\right)\nonumber\\
  & = & 
   \frac{M_6^4\Delta\phi}{2} \int d^4 x L,
\end{eqnarray}
where
%
\begin{equation}
 L =
-\eta^{\mu\nu}\partial_{\mu}\tilde{\bf Q}^T{\bf \Omega}
	  \partial_{\nu}\tilde{\bf Q} 
	  + \tilde{\bf Q}^T{\bf L}\tilde{\bf Q}
\end{equation}
Here, ${\bf L}$ and ${\bf \Omega}$ are defined in
(\ref{eqn:def-LOmega}) and
%
\begin{equation}
 \tilde{\bf Q} =
  \left(\begin{array}{c}
        \tilde{Q}_1[\tilde{\Phi}] \\ 
	\tilde{Q}_2[\tilde{\Phi}]
	\end{array} \right),
  \end{equation}
where
%
\begin{eqnarray}
 \tilde{Q}_1[\tilde{\Phi}] & \equiv & \tilde{\Phi}_2
  + \frac{2f\tilde{\Phi}_1}{6f+rf'}, \nonumber\\
 \tilde{Q}_2[\tilde{\Phi}] & \equiv & 
  \frac{1}{A'}\left[\frac{1}{2r^2}(fr^2\tilde{\Phi}_1)'+ f'\tilde{\Phi}_2
	      \right] 
  + \frac{frA'\tilde{\Phi}_1}{6f+rf'},
\end{eqnarray}
We have not written down the boundary term since it does not change the
definition of the generalized Klein-Gordon norm. It is easy to check
that the correct equations of motion are derived from this action.

From this form of the action we can read off the generalized
Klein-Gordon norm as
%
\begin{eqnarray}
 (\tilde{\Phi},\tilde{\Psi})_{KG} & \equiv & -i M_6^4\Delta\phi
  \int d^3{\bf x}\int_{r_-}^{r_+} dr r^2 \nonumber\\
 & &  \times
  \left[3\left(\tilde{Q}_1[\tilde{\Phi}]
	   \partial_t\tilde{Q}_1^*[\tilde{\Psi}]
	   -\tilde{Q}_1^*[\tilde{\Psi}]
	   \partial_t\tilde{Q}_1[\tilde{\Phi}]\right)
  + \frac{1}{f}
  \left(\tilde{Q}_2[\tilde{\Phi}]\partial_t\tilde{Q}_2^*[\tilde{\Psi}]
    -\tilde{Q}_2^*[\tilde{\Psi}]\partial_t\tilde{Q}_2[\tilde{\Phi}]\right)
  \right],
\end{eqnarray}
where the solutions $\tilde{\Phi}$ and $\tilde{\Psi}$ are specified by
pairs of five-dimensional functions ($\tilde{\Phi}_1$, $\tilde{\Phi}_2$)
and ($\tilde{\Psi}_1$, $\tilde{\Psi}_2$), respectively. For
$\tilde{\Phi}_i=\Phi_i(r) e^{i\eta_{\mu\nu}k_1^{\mu}x^{\nu}}$ and 
$\tilde{\Psi}_i=\Psi_i(r) e^{i\eta_{\mu\nu}k_2^{\mu}x^{\nu}}$, by using
the equations of motion, integrating by part and using $f(r_{\pm})=0$,
we obtain 
%
\begin{eqnarray}
 (\tilde{\Phi},\tilde{\Psi})_{KG}
  & \equiv & 
  M_6^4\Delta\phi (2\pi)^3(\omega_1+\omega_2)
  \delta^3({\bf k}_1-{\bf k}_2)e^{-i(\omega_1-\omega_2)t}
  \left\{
  \int_{r_-}^{r_+} 
  \frac{dr}{8(2rf'+6f+\Lambda_6r^2)} \right. \nonumber\\
 & &  \times 
  \left[
   \frac{1}{2}(m_1^2+m_2^2)r^2\Phi_1\Psi_1^*
   + 2r^4f'(\Psi_2^*\partial_r\Phi_1+\Phi_2\partial_r\Psi_1^*)
   \right.   \nonumber\\
 & & 
   + r^2(13f+3\Lambda_6r^2)\Phi_1\Psi_1^*
   + 4r^2(2rf'+12f+3\Lambda_6r^2)\Phi_2\Psi_2^*
   \nonumber\\
 & & \left.
   + \frac{r^2}{f}(r^2{f'}^2+10rff'+24f^2+3\Lambda_6r^2f)
   (\Phi_1+2\Phi_2)(\Psi_1^*+2\Psi_2^*) \right]   \nonumber\\
 & & 
  \left.
  + \left[\frac{3r^3 f'\Phi_1\Psi_1^*}
   {16(2f'+\Lambda_6r)}
  \right]_{r=r_-}^{r=r_+}\right\}, 
\label{KG_scalar}
\end{eqnarray}
where $\omega_i=k_i^0$ and $m_i^2=-\eta_{\mu\nu}k_i^{\mu}k_i^{\nu}$. It
is shown that $(\tilde{\Phi},\tilde{\Psi})_{KG}=0$ for $m_1^2\ne m_2^2$
and that $(\tilde{\Phi},\tilde{\Psi})_{KG}$ is time independent.


\end{document}